\documentclass[fleqn,usenatbib]{mnras}
\usepackage{newtxtext,newtxmath}

\usepackage[T1]{fontenc}
\usepackage{ae,aecompl}
\usepackage{macros}

\usepackage{geometry}
\usepackage{fancyhdr}
\usepackage{amsmath,amssymb}
\usepackage{graphicx} 
\usepackage[caption=false]{subfig}
\usepackage{hyperref}
\usepackage{wasysym}
\usepackage{lastpage}
\usepackage{etoolbox}
\usepackage{threeparttable}
\usepackage{multicol}

\graphicspath{{./}{Figures/}}

\title[Satellite orbital histories]{Orbital dynamics and histories of satellite galaxies around Milky Way-mass galaxies in the FIRE simulations}

\author[Santistevan et al.]{
Isaiah B. Santistevan$^{1}$\thanks{E-mail: ibsantistevan@ucdavis.edu},
Andrew Wetzel$^{1}$,
Erik Tollerud$^{2}$,
Robyn E. Sanderson$^{3,4}$,
Jenna Samuel$^5$
\\
$^{1}$Department of Physics \& Astronomy, University of California, Davis, CA 95616, USA\\
$^{2}$Space Telescope Science Institute, 3700 San Martin Dr, Baltimore, MD 21218, USA\\
$^{3}$Department of Physics \& Astronomy, University of Pennsylvania, Philadelphia, PA 19104, USA\\
$^{4}$Center for Computational Astrophysics, Flatiron Institute, New York, NY 10010, USA\\
$^{5}$Department of Astronomy, The University of Texas at Austin, 2515 Speedway, Stop C1400, Austin, TX 78712, USA\\
}

\date{Accepted XXX. Received YYY; in original form ZZZ}
\pubyear{2022}

\begin{document}
\label{firstpage}
\pagerange{\pageref{firstpage}--\pageref{lastpage}}
\maketitle

\begin{abstract}
The orbits of satellite galaxies encode rich information about their histories.
We investigate the orbital dynamics and histories of satellite galaxies around Milky Way (MW)-mass host galaxies using the FIRE-2 cosmological simulations, which, as previous works have shown, produce satellite mass functions and spatial distributions that broadly agree with observations.
We first examine trends in orbital dynamics at $z=0$, including total velocity, specific angular momentum, and specific total energy: the time of infall into the MW-mass halo primarily determines these orbital properties.
We then examine orbital histories, focusing on the lookback time of first infall into a host halo and pericentre distances, times, and counts.
Roughly 37 per cent of galaxies with $\Mstar\lesssim10^7\Msun$ were `pre-processed' as a satellite in a lower-mass group, typically $\approx2.7\Gyr$ before falling into the MW-mass halo.
Half of all satellites at $z=0$ experienced multiple pericentres about their MW-mass host.
Remarkably, for most (67 per cent) of these satellites, their most recent pericentre was not their minimum pericentre: the minimum typically was $\sim40$ per cent smaller and occurred $\sim6\Gyr$ earlier.
These satellites with growing pericentres appear to have multiple origins: for about half, their specific angular momentum gradually increased over time, while for the other half, most rapidly increased near their first apocentre, suggesting that a combination of a time-dependent MW-mass halo potential and dynamical perturbations in the outer halo caused these satellites' pericentres to grow.
Our results highlight the limitations of idealized, static orbit modeling, especially for pericentre histories.
\end{abstract}



\section{Introduction} %
\label{sec:intro}      %

The satellite galaxies around the Milky Way (MW) and M31 in the Local Group (LG) are unique systems in that we can measure both resolved stellar populations and full orbits to derive orbital histories.
Key answerable questions regarding satellite formation histories include:
What are their orbits today and what were their orbital histories?
When did they first become satellites?
How many were satellites of a lower-mass group, like the Large Magellanic Cloud \citep[LMC;][]{Kallivayalil18, Patel20, Pardy20}, and when did they become satellites of such groups?
How close have they orbited to the MW and M31?
Understanding the answers to these questions within the LG will also improve our understanding of satellite evolution in systems beyond the LG.

Thanks to numerous studies \citep[for example][]{Kallivayalil13, Kallivayalil18, Fritz18_segue, GaiaDR2, Patel20} and HST treasury programs (for example, GO-14734, PI Kallivayalil; GO-15902, PI Weisz), we now have or soon will have 3D velocity information for the majority of the satellites in the LG, with continued higher precision \citep[for example, see updated values between][and references therein]{McConnachie12, Fritz18}.
With the advent of new data coming from studies such as the Satellites Around Galactic Analogs (SAGA) survey \citep{SAGA_I, SAGA_II}, which focus on measuring properties of satellites of MW-mass galaxies beyond the LG, understanding general orbital/infall histories of satellites is imperative.
Given this rich data, combined with observationally informed estimates of the MW's gravitational potential \citep[for example][]{BlandHawthorn16, McMillan17, Li20, Deason21, CorreaMagnus22}, one can constrain the orbital histories of satellites.
However, a key challenge is understanding limitations and biases from assuming a static, non-evolving potential \citep[for example][]{Klypin99, DSouza22}.

One of the most important events in a satellite galaxy's history is when it first orbited within the MW-mass halo, or any more massive halo.
After falling into a more massive host halo, a satellite galaxy temporarily can orbit outside of the host's virial radius (a `splashback' satellite); typically such satellites remain gravitationally bound and orbit back within the host's virial radius \citep[for example][]{Ludlow09, Wetzel14}.
As a satellite orbits within the host halo, its hot gas can quench the satellite's star formation through ram-pressure stripping of gas \citep[for example][]{Gunn72, vandenBosch08, Fillingham19, RodriguezWimberly19, Samuel22}.
Not only can a lower-mass galaxy fall into a MW-mass halo, but it can become a satellite of another intermediate-mass galaxy before falling into the MW-mass halo, called `pre-processing' \citep[for example][]{DOnghia08, Li08, Wetzel15, Deason15, Kallivayalil18, Jahn19, Patel20}.
This process can suppress and even quench star formation in the low-mass galaxy before it falls into its MW-mass halo \citep[for example][]{Jahn21,Samuel22}.
If satellites fell into a MW-mass halo via a group, they should have similar orbits, at least for one or a few orbital timescales, with broadly similar orbital angular momenta and energy \citep[for example][]{Sales11, Deason15, Sales17, Pardy20}.
Some theoretical studies suggest that no current satellites of the MW were satellites during the epoch of reionization \citep[for example][]{Wetzel15, RodriguezWimberly19} at $z \gtrsim 6$, thus, if the satellites quenched at $z \gtrsim 6$, the host environment could not have quenched them, such that the effects of the host environment and cosmic reionization are separable, in principle.

Many works have studied infall histories and orbital properties of simulated satellite galaxies of MW-mass halos \citep[for example][]{Slater13, Wetzel15, Li20_infall, Bakels21, DSouza21, Robles21, Ogiya21}, sometimes with the intent of deriving properties of observed satellites of the MW \citep[for example][]{Rocha12, Fillingham19, Miyoshi20, RodriguezWimberly21}.
However, many such previous works used dark matter only (DMO) simulations, and the inclusion of baryonic physics is critical to model accurately the satellite population \citep[see][]{Brooks14, Bullock17, Sales22}.

One important process that affects the orbital evolution of satellites is dynamical friction.
As satellites orbit within a host halo, they induce an over-density of dark matter behind them, which causes a drag force called dynamical friction that slows their orbits and can lead them to merge with the host galaxy \citep{Chandrasekhar43, Ostriker75}.
Dynamical friction is more efficient when the satellite and host galaxy or halo are of similar mass \citep[for example][]{BoylanKolchin08, Jiang08, Wetzel10}.
Upon accretion, massive satellites can also induce global perturbations within the larger MW-mass galaxy, which also can affect the orbits of less-massive satellites \citep[for example][]{Weinberg86, Weinberg89, Colpi99, Tamfal21}.
Furthermore, because the dark matter in the satellite galaxy is tidally stripped as it orbits throughout the host halo, this stripped material further can slow the satellite \citep{Miller20}.
One way to parameterize the efficiency of dynamical friction is via the time from first infall it takes a satellite to merge into its host galaxy/halo.
For example, \citet{Wetzel10} approximated this merging time as 
\begin{equation}
\label{eq:dyn}
t_{\rm merge} \approx C_{\rm dyn} \frac{M_{\rm host}/M_{\rm sat}}{\ln(1+M_{\rm host}/M_{\rm sat})} t_{\rm Hubble}
\end{equation}
where $M_{\rm host}$ is the total mass of the host halo, $M_{\rm sat}$ is the total mass of the smaller satellite halo, $t_{\rm Hubble} = H^{-1}(z)$.
They found $C_{\rm dyn} \approx 0.2 - 0.3$ agrees well with the results of a large-volume cosmological DMO simulation.
This implies that for a satellite to merge within a Hubble time (the age of the Universe) the halo mass ratio between the host and a satellite must be closer than about $1:20$.

Because dynamical friction robs (primarily higher-mass) satellites of their orbital energy, one might expect satellites to shrink monotonically in orbital radius over time, until the satellite merges/disrupts.
Many studies implement idealized simulations (both with and without baryonic physics) that incorporate the effects of dynamical friction, mass loss, and ram-pressure stripping \citep[for example][]{Weinberg86, Taylor01, Penarrubia02, Penarrubia05, Amorisco17, Jiang21}, but testing this assumption in a cosmological setting is imperative.

Additionally, because the LMC has satellites of its own \citep[for example][]{DOnghia08, Deason15, Kallivayalil18}, many studies have investigated satellite spatial distributions and their dynamics with an additional LMC-like contribution to the host potential to test the dynamical effects of a massive satellite on nearby lower-mass galaxies \citep[for example][]{GaravitoCamargo19, Patel20, Samuel21, Vasiliev21, DSouza22, Pace22}.

In this paper, we examine the orbital histories of satellite galaxies using baryonic simulations that match general observed properties of satellites of MW-mass galaxies.
Specifically, we use the FIRE-2 cosmological zoom-in simulations, which form realistic MW-mass galaxies \citep[for example][]{GarrisonKimmel18, Hopkins18, Sanderson20, Bellardini21} and populations of satellite galaxies around them \citep[][]{Wetzel16, GarrisonKimmel19a, Samuel20, Samuel22}.
The two main topics that we explore are:
(1) The relation of orbital properties of satellite galaxies at $z = 0$ to their orbital histories, including lookback times of infall, distances from the MW-mass host, and stellar masses.
These relations not only help characterize the orbits of satellites, but their orbital histories also provide insight, or caution, into approximating history-based properties, such as infall time or pericentre information, based on their present-day properties, such as distance and stellar mass.
(2) Testing a common expectation that the orbits of satellite galaxies shrink over time, that is, that a satellite's most recent pericentric distance is the minimum that it has experienced.

In Santistevan et al., in prep, we will compare directly the orbits of satellites in cosmological simulations of MW-mass galaxies to orbits from integration in a static, idealized MW-mass halo \citep[see also ][and references therein]{Vasiliev21, DSouza22}.


\section{Methods}   %
\label{sec:methods} %

\subsection{FIRE-2 Simulations} %
\label{sec:sims}                %

\begin{table*}
\centering
\begin{threeparttable}
\caption{
Properties at $z = 0$ of the 13 MW/M31-mass galaxies in the FIRE-2 simulation suite that we analyze, ordered by decreasing stellar mass.
Simulations with `m12' names are isolated galaxies from the Latte suite, while the others are from the `ELVIS on FIRE' suite of LG-like paired hosts.
Columns: host name; $M_{\rm star,90}$ is the host's stellar mass within $R_{\rm star,90}$, the disk radius enclosing 90 per cent of the stellar mass within 20 kpc; $M_{\rm 200m}$ is the halo total mass; $R_{\rm 200m}$ is the halo radius; $N_{\rm satellite}$ is the number of satellite galaxies at $z=0$ with $\Mstar>3\times10^4\Msun$ that ever orbited within $R_{\rm 200m}$; $M_{\rm sat,star}^{\rm max}$ is the stellar mass of the most massive satellite at $z = 0$; and $M_{\rm sat,halo,peak}^{\rm max}$ is the peak halo mass of the most massive satellite.
In Remus and Juliet, the satellite with the largest stellar mass is not the same as the satellite with the largest subhalo mass.
}
\begin{tabular}{|c|c|c|c|c|c|c|c|c|}
		\hline
		\hline
		Name & $M_{\rm star,90}$ & $M_{\rm 200m}$ & $R_{\rm 200m}$ & $N_{\rm satellite}$ & $M^{\rm sat}_{\rm star,max}$ & $M^{\rm sat}_{\rm halo,peak,max}$ & Reference \\
		 & [$10^{10} \Msun$] & [$10^{12} \Msun$] & [kpc] & & [$10^8 \Msun$] & [$10^{10} \Msun$] &  \\
		\hline
        m12m    & 10.0 & 1.6 & 371 & 44 & 4.68     & 3.78     & A \\
        Romulus & 8.0  & 2.1 & 406 & 52 & 2.34     & 2.68     & B \\
        m12b    & 7.3  & 1.4 & 358 & 32 & 0.58     & 1.21     & C \\
        m12f    & 6.9  & 1.7 & 380 & 43 & 1.61     & 2.01     & D \\
        Thelma  & 6.3  & 1.4 & 358 & 33 & 0.33     & 1.78     & C \\
        Romeo   & 5.9  & 1.3 & 341 & 33 & 1.92     & 3.34     & C \\
        m12i    & 5.5  & 1.2 & 336 & 26 & 1.24     & 2.40     & E \\
        m12c    & 5.1  & 1.4 & 351 & 40 & 15.1     & 16.7     & C \\
        m12w    & 4.8  & 1.1 & 319 & 38 & 7.79     & 4.88     & F \\
        Remus   & 4.0  & 1.2 & 339 & 34 & 0.50$^*$ & 2.07$^*$ & B \\
        Juliet  & 3.3  & 1.1 & 321 & 38 & 2.51$^*$ & 3.16$^*$ & C \\
        Louise  & 2.3  & 1.2 & 333 & 34 & 2.50     & 3.71     & C \\
        m12r    & 1.5  & 1.1 & 321 & 27 & 28.4     & 13.7     & F \\
		\hline
		\hline
\end{tabular}
\label{tab:hosts}
\begin{tablenotes}
\item \textit{Note:} Simulation introduced in: A: \citet{Hopkins18}, B: \citet{GarrisonKimmel19b}, C: \citet{GarrisonKimmel19a}, D: \citet{GarrisonKimmel17}, E: \citet{Wetzel16}, F: \citet{Samuel20}.
\end{tablenotes}
\end{threeparttable}
\end{table*}

We use the cosmological zoom-in baryonic simulations of both isolated MW/M31-mass galaxies and LG-like pairs from the Feedback In Realistic Environments (FIRE) project\footnote{See the FIRE project web site: http://fire.northwestern.edu} \citep{Hopkins18}.
We ran these simulations using the hydrodynamic plus $N$-body code \textsc{Gizmo} \citep{Hopkins15}, with the mesh-free finite-mass (MFM) hydrodynamics method \citep{Hopkins15}.
We used the FIRE-2 physics model \citep{Hopkins18} that includes several radiative heating and cooling processes such as Compton scattering, Bremsstrahlung emission, photoionization and recombination, photoelectric, metal-line, molecular, fine-structure, dust-collisional, and cosmic-ray heating across temperatures $10-10^{10}\K$, including the spatially uniform and redshift-dependent cosmic ultraviolet (UV) background from \cite{FaucherGiguere09}, for which HI reionization occurs at $z_{\rm reion} \approx 10$.
Star formation occurs in gas that is self-gravitating, Jeans unstable, molecular \citep[following][]{Krumholz11}, and dense ($n_{\rm H} > 1000$ cm$^{-3}$).
Star particles represent single stellar populations, assuming a \cite{Kroupa01} initial mass function, and they evolve along stellar population models from \textsc{STARBURST99 v7.0} \citep{Leitherer99}, inheriting masses and elemental abundances from their progenitor gas cells.
FIRE-2 simulations also include the following stellar feedback processes: core-collapse and Ia supernovae, stellar winds, and radiation pressure.

We used the code \textsc{MUSIC} \citep{Hahn11} to generate the cosmological zoom-in initial conditions at $z \approx 99$ 
within periodic cosmological boxes of length $70.4 - 172 \Mpc$, sufficiently large to avoid unrealistic periodic gravity effects on individual MW-mass halos.
For each simulation, we save 600 snapshots with $20 - 25 \Myr$ spacing down to $z=0$, assuming a flat $\Lambda$CDM cosmology.
Consistent with \citet{Planck18}, we used cosmological parameters in the following ranges: $h = 0.68 - 0.71$, $\sigma_{\rm 8} = 0.801 - 0.82$, $n_{\rm s} = 0.961 - 0.97$, $\Omega_{\Lambda} = 0.69 - 0.734$, $\Omega_{\rm m} = 0.266 - 0.31$, and $\Omega_{\rm b} = 0.0449 - 0.048$.

Our galaxy sample consists of the 12 MW/M31-mass galaxies in \citet{Santistevan20}, as well as one additional galaxy, `m12r' first introduced in \citet{Samuel20}.
These are from the Latte suite of isolated MW/M31-mass galaxies introduced in \citet{Wetzel16} and the `ELVIS on FIRE' suite of LG-like MW+M31 pairs, introduced in \citet{GarrisonKimmel19a}.
Table~\ref{tab:hosts} lists their stellar mass, $M_{\rm star,90}$, halo mass, $M_{\rm 200m}$, and radius $R_{\rm 200m}$ at $z = 0$, where both are defined at 200 times the matter density at $z = 0$.
The Latte suite consists of halos with $M_{\rm 200m} = 1 - 2 \times 10^{12} \Msun$ at $z = 0$ with no other similar-mass halos within $5 \times R_{\rm 200m}$.
We also chose m12r and m12w to have LMC-mass satellite analogs near $z \approx 0$ \citep{Samuel20}.
The initial masses of star particles and gas cells is $7100 \Msun$, but the average star particle mass at $z = 0$ is $\approx 5000 \Msun$ from stellar mass loss.
Within the zoom-in region, the mass of DM particles is $3.5 \times 10^4 \Msun$.
The gravitational softening lengths are 4 and 40 pc (Plummer equivalent), co-moving at $z > 9$ and physical thereafter, for star and DM particles, respectively.
Gas cells use adaptive force softening, equal to their hydrodynamic smoothing, down to 1 pc.

Each pair of halos in the `ELVIS on FIRE' suite of LG-like pairs was chosen based on their individual mass ($M_{\rm 200m} = 1 - 3 \times 10^{12} \Msun$) and combined masses (total LG mass between $2 - 5 \times 10^{12} \Msun$), as well as their current separation ($600 - 1000\kpc$) and radial velocities at $z = 0$ ($\rm \upsilon_{rad} < 0$), and isolated environment (no other massive halos within $2.8 \Mpc$ of either host center).
The mass resolution is $\approx 2 \times$ better in the `ELVIS on FIRE' suite, with initial baryonic particle masses $3500 - 4000 \Msun$.
For all results in this paper, we investigated possible differences between the isolated and LG-like MW-mass galaxies, which also partially tests resolution convergence, and we find negligible differences between the two samples.

These 13 host galaxies reflect formation histories of general MW/M31-mass (or LG-like) galaxies within our selection criteria and exhibit observational properties broadly similar to the MW and M31, including:
realistic stellar halos \citep{Bonaca17, Sanderson18}, dynamics of metal-poor stars from early galaxy mergers \citep{Santistevan21}, satellite galaxy stellar masses and internal velocity dispersions \citep{Wetzel16, GarrisonKimmel19b}, radial and 3-D spatial distributions \citep{Samuel20, Samuel21}, and star-formation histories and quiescent fractions \citep{GarrisonKimmel19b, Samuel22}.

\subsection{Halo/Galaxy Catalogs and Merger Trees} %
\label{sec:rockstar}                               %

We use the \textsc{ROCKSTAR} 6-D halo finder \citep{Behroozi13a} to generate (sub)halo catalogs using only DM particles at each of the 600 snapshots, and \textsc{CONSISTENT-TREES} \citep{Behroozi13b} to generate merger trees.
As a consequence of the large zoom-in volume for each host, there is no low-resolution DM particle contamination in any of the (sub)halos that we analyze.

\citet{Samuel20} describes our star particle assignment to (sub)halos in post-processing; we briefly review it here.
We first select star particles within $d < 0.8 R_{\rm halo}$ (out to a maximum distance of $30 \kpc$) with velocities $\upsilon < 2 V_{\rm circ,max}$ of the (sub)halo's center-of-mass (COM) velocity.
We then keep only the star particles within $d < 1.5 R_{\rm star,90}$ (the radius enclosing 90 per cent of the stellar mass) of the (then) current member stellar population's COM and halo center position.
We further kinematically select the star particles with velocities $\upsilon < 2 \sigma_{\rm vel,star}$ (the velocity dispersion of current member star particles) of the COM velocity of member star particles.
Finally, we iterate on these spatial and kinematic criteria, which guarantees that the COM of the galaxy and (sub)halo are consistent with one another, until the (sub)halo's stellar mass converges to within 1 per cent.

We use two publicly available analysis packages: \textsc{HaloAnalysis}\footnote{\url{https://bitbucket.org/awetzel/halo\_analysis}} \citep{HaloAnalysis} for assigning star particles to halos and for reading and analyzing halo catalogs/trees, and  \textsc{GizmoAnalysis}\footnote{\url{https://bitbucket.org/awetzel/gizmo\_analysis}} \citep{GizmoAnalysis} for reading and analyzing particles from Gizmo snapshots.

\subsection{Selection of Satellites}
\label{sec:selection}

We include all luminous satellite galaxies at $z = 0$ with $\Mstar > 3\times10^4\Msun$ that have crossed within their MW-mass host halo's $R_{\rm 200m}(z)$.
This lower limit on stellar mass corresponds to $\approx 6$ star particles, the limit for reasonably resolving the total stellar mass \citep{Hopkins18}.
Our sample includes `splashback' satellites that are currently beyond the host's $R_{\rm 200m}$, which are typically still gravitationally bound to the host but simply near apocentre \citep[for example][]{Wetzel14}.
As Table~\ref{tab:hosts} shows, the number of surviving luminous satellites at $z = 0$, including this splashback population, per host ranges from 26-52, and our sample totals 473 satellites.
Both \citet{GarrisonKimmel14} and \citet{Wetzel15} showed that in the ELVIS DMO simulation suite the average number of subhalos that would typically host galaxies with $\Mstar\gtrsim10^5\Msun$ is $\sim31-45$.
However, because stellar feedback and baryonic physics can affect galaxy formation, \citet{Wetzel16} and \citet{GarrisonKimmel19a} showed that the number of satellites above this mass range decreases to $\sim13-15$.
More recently, \citet{Samuel20} showed that the radial distributions of these satellites are consistent with the MW and M31 out to $300\kpc$.
The MW and M31 each have 13 and 27 satellites, respectively, and the MW-mass hosts in our simulations bracket these values with 11 to 27 satellites.
Unless otherwise stated, in our analysis we refer to luminous satellites, i.e. satellites containing stars, as simply `satellites'.

In computing host-averaged results below, to avoid biasing our results to the hosts with larger satellite populations, we oversample the satellites so that each host contributes a nearly equal fraction of satellites to the total.
Specifically, we multiply the number of satellites in the MW-mass host with the largest population (Romulus, with 52 satellites) by 10 which results in an oversampled population of 520.
Then, for each of the other MW-mass hosts, we divide 520 by the number of their satellites and obtain the nearest integer multiplicative factor, $m$, that we apply to each host's satellite population, $N_{\rm sat}$ (see Table~\ref{tab:hosts}), so that each host contains $\approx 500 - 530$ satellites, or that their satellite populations are within 5 per cent of one another.
Thus, when plotting properties, such as pericentre distances, we count each satellite in a given host $m$ times for each property in the figures.

\begin{figure}
\centering
\begin{tabular}{c}
\includegraphics[width=0.94\linewidth]{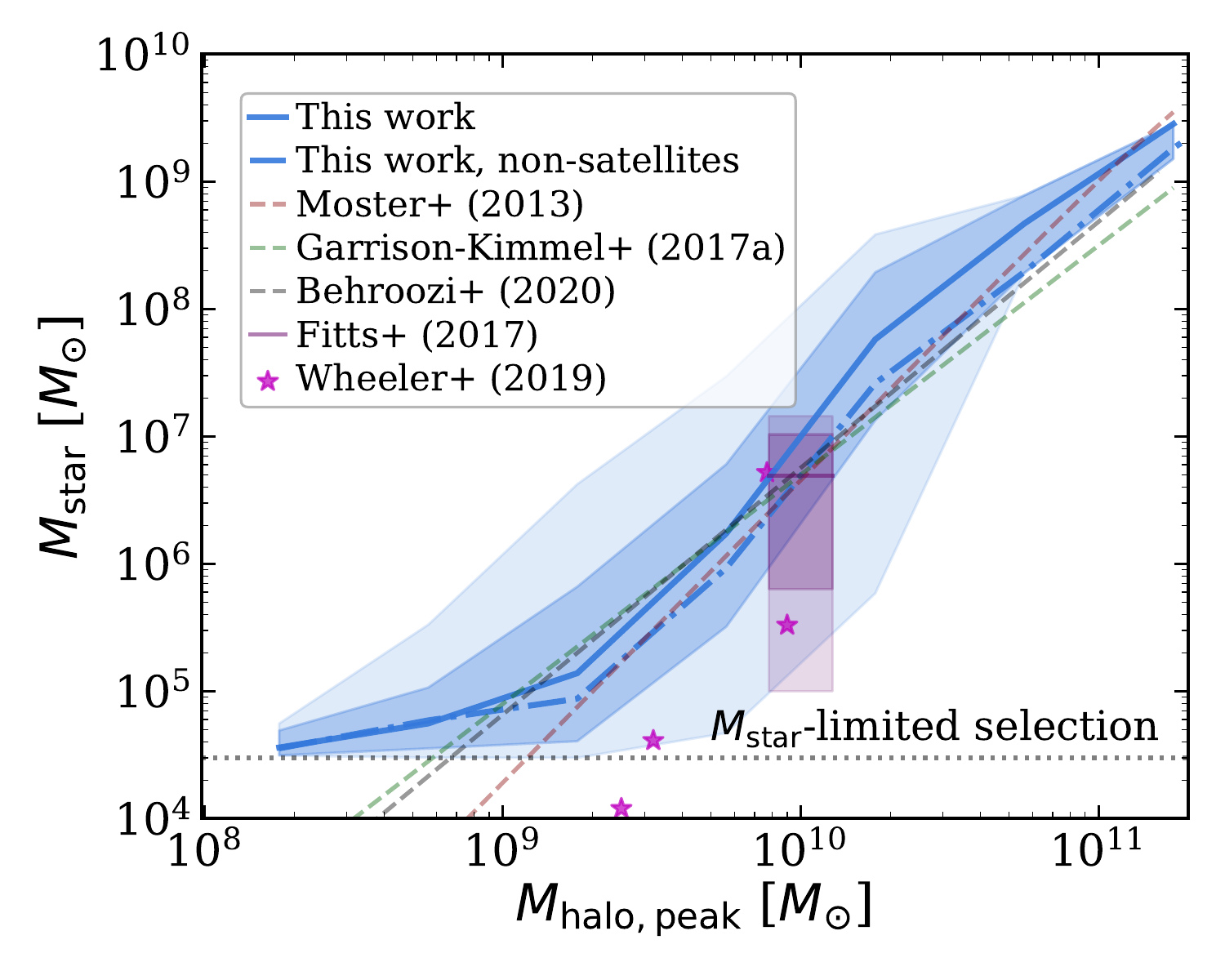}
\end{tabular}
\vspace{-2 mm}
\caption{
Stellar mass, $\Mstar$, at $z = 0$ versus peak halo mass, $\Mhp$, for all 473 satellites with $\Mstar > 3 \times 10^4 \Msun$ across all 13 host galaxies.
The solid blue line shows the median, and the dark and light shaded regions show the 68th percentile and full distribution, respectively.
We compare this with the stellar-halo mass relation of non-satellite low-mass galaxies which never fell into the MW-mass host, and currently orbit beyond $1\Mpc$ at $z = 0$ (blue dot-dashed).
The dotted grey line indicates the minimum $\Mstar$ in our sample.
For comparison, we also show extrapolations of the stellar-halo mass relations from \citet{Moster13} (red), \citet{GarrisonKimmel17_smhm} (green), and \citet{Behroozi20} (black), which broadly agree with our sample at $\Mhp\gtrsim10^9\Msun$.
Additionally, we show the 68th percentile and full distribution of values from higher resolution, \textit{isolated} low-mass galaxies from \citet{Fitts17} in the dark and light shaded pink regions, respectively, and the median across their sample in the horizontal pink line.
We finally show 4 higher resolution, isolated low-mass galaxies from \citet{Wheeler19} as stars, however, the galaxy with the smallest $\Mstar$ in this sample is beyond our resolution limit.
At all halo masses, the 68th percentile in satellite $\Mstar$ is $0.5 - 1$ dex, with a smaller range at the extreme masses from lower statistics.
The relation flattens at $\Mstar \lesssim 10^5 \Msun$ simply because of our stellar mass limit; \textbf{therefore, our fiducial sample is complete in $\Mstar$ to $> 3 \times 10^4 \Msun$, while we are complete in $\Mhp$ to $\gtrsim 3 \times 10^9 \Msun$} (see Appendix~\ref{app:mhalo} for results for selecting satellites via $\Mhp$).
}
\label{fig:mstar_v_mhalo}
\end{figure}

Figure~\ref{fig:mstar_v_mhalo} shows the relation between stellar mass and halo mass (SMHM) for our satellites.
We show stellar mass \textit{at $z = 0$} versus \textit{peak} (sub)halo mass throughout its history.
We include the median SMHM relation of non-satellites in the same simulations, which we define as low-mass galaxies that never crossed the virial radius of the MW-mass host, and that currently orbit beyond $1\Mpc$ at $z = 0$, in the blue dot-dashed line.
The formation histories of non-satellites differ from satellites because they form in less dense regions of the Universe.
This, along with the UV heating of gas in isolated low-mass galaxies may explain their slightly smaller $\Mstar$, however, a deeper investigation is outside of the scope of this paper.
The grey dotted line at $\Mstar = 3\times10^4\Msun$ shows our lower limit in stellar mass.

To compare to other SMHM relations from the literature, we also show extrapolations from \citet{Moster13, GarrisonKimmel17_smhm, Behroozi20} in the dashed red, green, and black lines, respectively.
We also include the median $\Mstar$, 68th percentile, and full distribution of values from \citet{Fitts17} of higher resolution (initial baryon masses of $500\Msun$), isolated low-mass galaxies, in the pink horizontal line, and dark and light pink shaded regions, respectively.
Finally, we include 4 higher resolution, isolated low-mass galaxies from \citet{Wheeler19} (named m09, m10q, m10v, and m10vB in their work), with initial baryon masses of $30\Msun$ (magenta stars).
The SMHM relation in our sample broadly agrees with these (extrapolated) semi-empirical estimates and values of isolated low-mass galaxies; for a more detailed discussion about the SMHM relation in FIRE-2, see \citet{Hopkins18, Wheeler19}.
The low-mass end of our SMHM relation in Figure~\ref{fig:mstar_v_mhalo} flattens at $\Mstar\lesssim10^5\Msun$, purely because of our stellar mass selection of $\Mstar>3\times10^4\Msun$.
We note that the galaxy with the smallest $\Mstar$ from \citet{Wheeler19} is beyond the resolution limit in our sample, and the second smallest $\Mstar$ galaxy would be only marginally resolved.
However, with better resolution, the SMHM relation in Figure~\ref{fig:mstar_v_mhalo}, would likely follow a similar trend as the extrapolations and it is likely that the lowest $\Mstar$ isolated galaxy would lie in the lower end of the full distribution scatter.
Thus, while our sample is complete in stellar mass, we are complete in halo mass for $\Mhp\gtrsim3\times10^9\Msun$.
We find only minor differences in our results if selecting satellites via $\Mhp$ instead (see Appendix~\ref{app:mhalo}).

We checked the SMHM relation in Figure~\ref{fig:mstar_v_mhalo} instead using the \textit{peak} stellar mass throughout a galaxy's history, $M_{\rm star,peak}$.
The two relations are similar, but the SMHM relation with $M_{\rm star,peak}$ has a $\approx 5\%$ higher normalization, on average, because of stellar mass loss after infall.

\subsection{Numerical Disruption} %
\label{sec:disrupt}               %

Many previous studies \citep[for example][]{vandenBosch18_I, vandenBosch18_II, Bovy20} noted that without proper mass and spatial resolution, satellite subhalos can suffer from artificial numerical disruption too quickly.
Thus, sufficient resolution and implementation of the relevant physics is necessary to model accurately the evolution of satellite galaxies.
As a partial test of this, we investigated differences between the isolated and LG-like satellite populations, which have a $\approx 2 \times$ resolution difference, and we saw no strong differences in our results.
\citet{Samuel20} also tested resolution convergence of the satellite radial profiles using the FIRE-2 simulations with dark matter particle masses of $m_{\rm dm} = 3.5 \times10^4\Msun$ and $m_{\rm dm} = 2.8 \times10^5\Msun$ and found generally consistent results between the two, because there are enough dark matter particles ($ \gtrsim 2 \times 10^4$) in the lowest-mass luminous subhalos ($\Mhp \gtrsim 10^8 \Msun$) to prevent numerical disruption, and many more than that in our typical (more massive) luminous subhalos, thus satisfying criteria such as in \citet{vandenBosch18_II}.
Also, our DM particle mass resolution is $m_{\rm dm} = 2 - 3.5\times10^4\Msun$ and DM force softening length is $40$ pc, significantly better than previous work like \citet{Wetzel15}, who used the ELVIS DMO simulations with $m_{\rm dm} = 1.9 \times 10^5 \Msun$ and $140$ pc but found results broadly consistent with ours.
Nonetheless, any simulations like ours necessarily operate at finite resolution, which inevitably leads to some degree of numerical disruption, which any reader should bear in mind.

\subsection{Calculating pericentre} %
\label{sec:pipeline}                %

We calculate pericentres by tracking the main progenitor back in time using the merger trees, which store each satellite's galactocentric distance at each of the 600 snapshots.
We first ensure that the satellite is within the MW-mass host halo at a given snapshot.
Then, we check if the galactocentric distance reaches a minimum within $\pm 20$ snapshots, corresponding to a time window of $\approx 1 \Gyr$.
Given the $\approx 25 \Myr$ time spacing between snapshots, we fit a cubic spline to the distance and time arrays across this interval.
We then find the minimum in the spline-interpolated distance(s), and record the corresponding spline-interpolated time.

We checked how our results differ if varying the window to $\pm$ 4, 8, and 10 snapshots by visually inspecting each satellite's orbit history and conclude that a window of $\pm 20$ snapshots reduces nearly all `false' pericentres, that is, instances in which the criteria above are met because of numerical noise in the orbit or a short-lived perturbation.
Because the center of mass of the MW-mass galaxy does not perfectly coincide with the center of mass of its DM host halo, we also checked how our results vary between using distances with respect to the center of the galaxy versus the halo: we find no meaningful difference.
We additionally checked how our results vary when using distances with respect to the center of the satellite galaxy versus the center of the satellite (sub)halo, again finding no significant differences.

\subsection{Calculating Gravitational Potential and Total Energy}
\label{sec:potential}

We also explore trends with a satellite's specific orbital total energy, $E$ at $z = 0$, defined as the sum of the kinetic and potential energies.
Our simulations store the value of the gravitational potential for all particles at each snapshot, so to calculate the potential for each satellite at $z = 0$, we select all star, gas, and DM particles within $\pm 5 \kpc$ of the satellite (sub)halo's virial radius, to limit biasing from the satellite's self-potential, and we compute the mean potential across these particles.
Given that some satellites are in LG-like environments, we normalize $E$ at the MW-mass halo radius, such that $E(d = R_{\rm 200m}) = 0$, that is, the sum of the host potential at $R_{\rm 200m}$ and the kinetic energy of a circular orbit at $R_{\rm 200m}$ is 0.


\section{Results}   %
\label{sec:results} %

Throughout the paper, we present results for all satellite galaxies at $z = 0$, based on their stellar mass ($\Mstar > 3 \times 10^4 \Msun$), across all of our MW-mass hosts.
Although \citet{Santistevan20} noted that MW-mass halos/galaxies in LG-like pairs formed $\sim 1.6 \Gyr$ earlier than those that are isolated, we compared satellites based on isolated versus LG-like environment and find negligible differences in any properties that we investigate.
This agrees with the lack of dependence in \citet{Wetzel15}, who investigated the infall histories of satellite subhalos in the ELVIS dark matter-only simulations.
Appendix~\ref{app:mhalo} examines how our results change by selecting satellites by their peak halo mass.
In summary we find qualitatively similar results for selecting via stellar mass, but given the scatter in the SMHM relation, the trends with halo mass are smoother and any features are sharper.
Although the dark matter (sub)halo mass is more dynamically relevant to the orbits of satellite galaxies, we present our results with a stellar mass selected sample, because it is observationally easier to measure than halo mass.
Finally, Appendix~\ref{app:dmo} compares our results for baryonic simulations against DMO simulations of the same systems.
In summary, the lack of a MW-mass galaxy in the DMO simulations allows satellites to survive longer, orbit closer to the center of the halo, and complete more orbits.

We examine trends guided by which orbital properties are relevant to different phenomena.
As we will show, specific angular momentum and specific total energy provides insight into when a satellite fell into the MW-mass halo.
We also explore trends with satellite mass, in part to understand where dynamical friction becomes important: from Equation~\ref{eq:dyn}, for the MW-mass halos with $M_{\rm 200m}(z = 0) \approx 10^{12}\Msun$ (and lower $M_{\rm 200m}$ at earlier times), we expect dynamical friction to significantly affect satellites with $M_{\rm 200m} \gtrsim 3 \times 10^{10}\Msun$, or $\Mstar \gtrsim 10^8 \Msun$.
We also focus on infall times, to understand how long satellites have been orbiting in the host halo environment, and we explore the incidence of pre-processing in a lower-mass group prior to MW-mass infall.
We also examine properties of orbital pericentre, given that satellites typically feel the strongest gravitational tidal force and the strongest ram pressure at pericentre.

In this paper, we present trends for the simulated satellite populations only, however, in the future we plan to investigate differences in the simulations and results obtained from idealized orbit modeling methods.
Ultimately we will provide a framework to derive similar orbital properties from satellites in the MW and M31 using the satellite populations in the simulations, and compare the results.
Thus, we leave direct observational comparisons for future work.

\subsection{Orbital properties today}
\label{sec:dynamics}

\begin{figure*}
\centering
\begin{tabular}{c}
\includegraphics[width=0.95\linewidth]{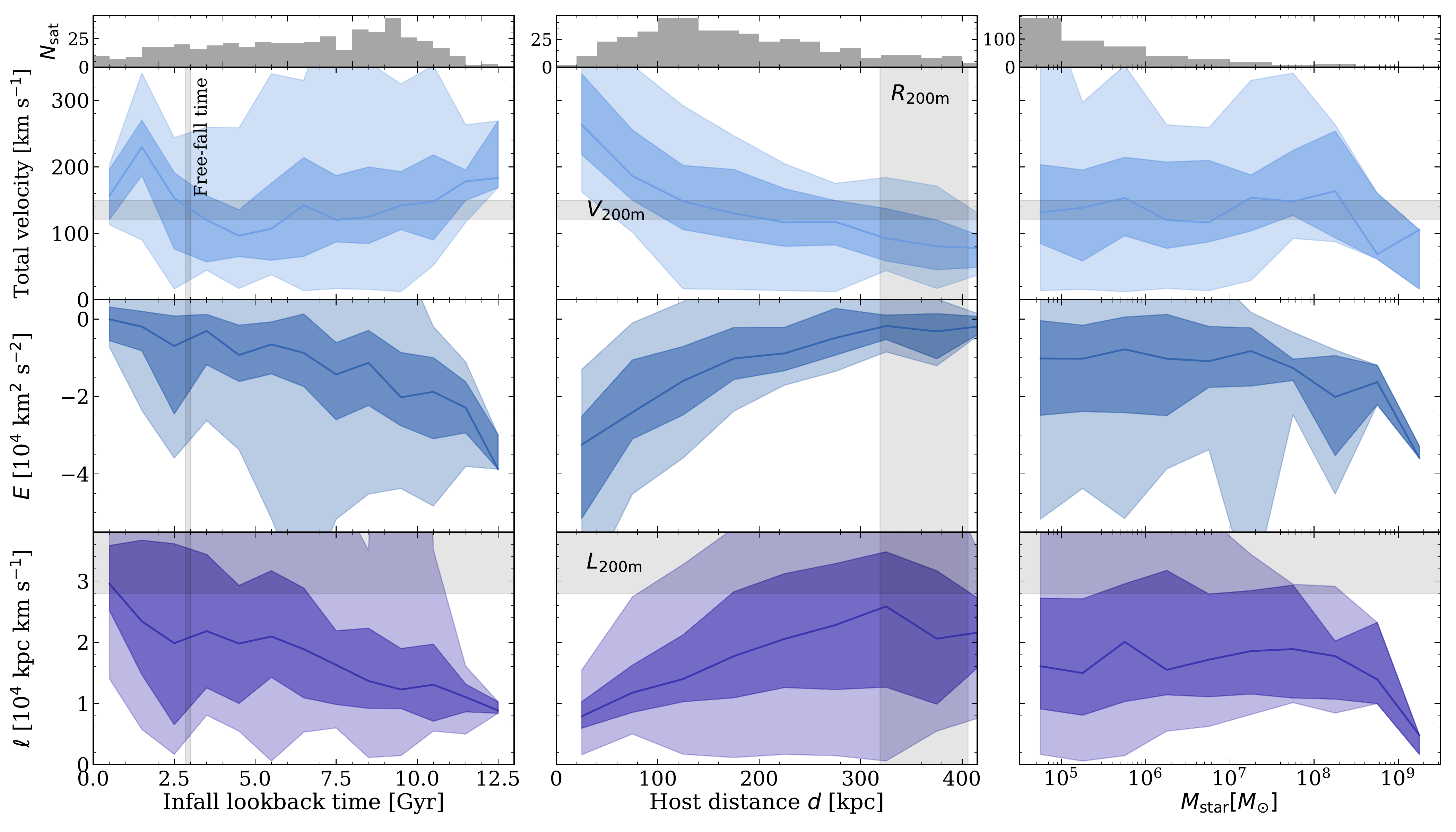}
\end{tabular}
\vspace{-2 mm}
\caption{
Instantaneous orbital dynamics of satellite galaxies at $z = 0$ versus their lookback time of infall into the MW-mass halo (left), distance from MW-mass host, $d$ (middle), and stellar mass, $\Mstar$ (right).
The solid lines show the median for all satellites across our 13 MW-mass hosts, and the dark and light shaded regions show the 68th percentile and full distribution, respectively.
At the top of each column, we show the histogram of these properties for the sample.
We show common reference points to these properties for typical host halos within our mass range in the horizontal and vertical gray shaded bands in some of these panels, such as the free-fall time, and virial radii, velocities, and angular momenta; see text for details.
\textbf{Top row}: Orbital total velocity.
The median velocity for satellites that fell in $>4 \Gyr$ ago increases with earlier infall time from $100 - 200 \kmsi$ (top left).
However, more recently infalling satellites show a peak near $\approx 1.5 \Gyr$ of $\approx 230 \kmsi$, highlighting satellites that are nearing their first pericentre, before decreasing again to $\approx 100 \kmsi$ for satellites that only recently fell in.
Velocity also decreases with host distance and is relatively flat with $\Mstar$ at roughly $135\kmsi$.
\textbf{Middle row}: Specific orbital total energy, $E$.
We normalize $E$ at the host's $R_{\rm 200m}$ so $E(d = R_{\rm 200m}) = 0$.
$E$ decreases for earlier-infalling satellites, with median values ranging from $0$ to $-3.8\times10^4\kmsis$.
Satellites that fell in earlier orbit deeper in the host halo gravitational potential and therefore are more bound (middle left).
Satellites with $\Mstar \gtrsim 10^8 \Msun$ are more bound, because of the stronger dynamical friction they experience, and satellites closer to the host are more bound, because they orbit deeper in the gravitational potential (middle right).
\textbf{Bottom row}: Orbital specific angular momentum, $\ell$.
Specific angular momentum decreases for earlier-infalling satellites, from $\approx 3$ to $0.9 \times 10^4 \kpc\kmsi$ (bottom left), because satellites that fell in earlier fell in (and orbit) at smaller distances (see Figure~\ref{fig:infall_dz0} and bottom left panel of Figure~\ref{fig:peri_dn}).
$\ell$ necessarily increases with $d$ (bottom middle) and depends only weakly on satellite $\Mstar$ (bottom right).
}
\label{fig:dynamics}
\end{figure*}

We first investigate the instantaneous orbital properties of satellites at $z = 0$, including approximate integrals of motion like angular momentum and energy.
Figure~\ref{fig:dynamics} shows total velocity, specific orbital total energy, $E$, and specific angular momentum, $\ell$, as a function of lookback time of infall into the MW-mass halo, $\MWinfall$, galactocentric distance, $d$, and stellar mass, $\Mstar$.
The top of each column shows distributions of $\MWinfall$, $d$, and $\Mstar$.
In particular, the distribution of $\MWinfall$ is relatively flat, with a modest peak $\approx 9 \Gyr$ ago.
For reference, all panels versus $\MWinfall$ (left) include a vertical shaded region to represent the free-fall time at $R_{\rm 200m}$, $t_{\rm ff} = \sqrt{ 3 \pi / \left(32 G \rho_{\rm 200m}\right) }$, where $t_{\rm ff} = 2.8 - 3 \Gyr$ across our hosts.
In all panels versus $d$ (middle), the vertical shaded region shows the range of $R_{\rm 200m}$ for the MW-mass halos, as Table~\ref{tab:hosts} lists.
Similarly, all panels versus total velocity and $\ell$ show horizontal shaded bands that represent the range of $V_{\rm 200m}$ and $L_{\rm 200m}$ across hosts, where $V_{\rm 200m} = \sqrt{GM_{\rm 200m}/R_{\rm 200m}}$ is the velocity of a circular orbit at the virial radius, and $L_{\rm 200m} = V_{\rm 200m}\times R_{\rm 200m}$ is the specific angular momentum of that orbit.
Because the satellites fell in at different times and orbit at various distances with unique stellar masses, we do not expect them to have values equal to $V_{\rm 200m}$ or $L_{\rm 200m}$, so we provide these shaded regions as a reference only.

\subsubsection{Total velocity} %

We first present trends in the total velocity, corresponding to the top row in Figure~\ref{fig:dynamics}.
Considering the trends in total velocity with infall time (top left), satellites that fell in $< 1 \Gyr$ ago have not yet experienced a pericentre, so they show an increase in total velocity with time since infall from $0.5 - 1.5 \Gyr$ ago, because these satellites are near pericentre.
The total velocity then decreases with increasing time since infall, because those satellites are now near apocentre.
We see a similar, but weaker, peak in total velocity $\sim 6 \Gyr$ ago, from a marginally phase-coherent population near pericentre, but after a few orbits, satellites become out of phase.
Satellites that fell in $\lesssim 3 \Gyr$ ago typically have total velocities of $150 - 250 \kmsi$, while earlier infalling satellites are only orbiting at $100 - 185 \kmsi$.
Comparing these infall times to $t_{\rm ff}$ in the vertical shaded bands, satellites with $\MWinfall < t_{\rm ff}$ often have larger velocities, and again are likely to be on first infall, because they have not had enough time to reach pericentre.
For reference, we also compare the satellite total velocities to the host's virial velocity, $V_{\rm 200m}$ (horizontal gray shaded band).
Satellites that fell in $\MWinfall \approx 3 - 11 \Gyr$ ago typically have comparable total velocities to the virial velocity, but the full scatter extends to both much larger and smaller values.

Next, the median velocity decreases with increasing $d$ (top middle), from as high as $260\kmsi$ for the closest satellites to $80\kmsi$ for satellites near $R_{\rm 200m}$.
The shaded region of the 68th percentile follows the median trend and the width is roughly constant at $\approx 95 \kmsi$ across all distances.
At $R_{\rm 200m}$, the total velocities of satellites are typically lower than $V_{\rm 200m}$ both because they are not on perfectly circular orbits and because the population at large $d$ is likely biased to lower values from the splashback population which typically have negligible velocities.

The median is nearly constant with $\Mstar$ (top right), ranging from $120 - 160 \kmsi$ for $\Mstar \approx 10^{4.75-8.25} \Msun$, which decreases to $70 - 100 \kmsi$ at higher mass, likely because sufficiently massive satellites experience significant dynamical friction that slows their orbit.
Across all stellar masses, the average total velocity is $\approx 135 \kmsi$ and the typical range of the 68th percentile is $90 - 205 \kmsi$.
The median for all satellites with $\Mstar \lesssim 10^{8.5}\Msun$ show consistent values with $V_{\rm 200m}$.

\subsubsection{Specific total energy} %

Next, the middle row in Figure~\ref{fig:dynamics} shows trends in the specific orbital energy, $E$.
Given that some satellites are in LG-like environments, which complicates computing the specific total energy beyond $R_{\rm 200m}$, we normalize $E$ at the MW-mass halo's $R_{\rm 200m}$ so $E(d = R_{\rm 200m}) = 0$.
Thus, satellites with $E > 0$ are the splashback population with apocentres beyond $R_{\rm 200m}$ and are essentially all bound to the host halo.

The middle left panel shows that earlier infalling satellites are on more bound orbits, with the median $E$ decreasing from $\approx 0$ to $-3.9 \times 10^4\kmsis$.
This reflects the growth of the MW-mass halo over time.
$E$ increases with $d$ from the host, so satellites are more bound at smaller distances.
Thus, the median $E$ is not constant with $d$, but this does not imply that specific energy is not being conserved over an orbit.
As we explore below, $\MWinfall$ correlates with $d$, such that satellites at smaller $d$ fell in earlier (though with large scatter; see Figure~\ref{fig:infall_dz0}).
This is largely because they fell in when the host $R_{\rm 200m}$ was smaller, so they necessarily orbit at smaller $d$.
This then leads to the correlation of $E$ with $d$ across the population at $z = 0$ (Figure~\ref{fig:dynamics}, center panel).

Similar to the trends in total velocity, $E$ does not strongly depend on $\Mstar$, except at $\Mstar\gtrsim10^8\Msun$, where satellites experience significant dynamical friction, causing their orbits to become more bound, despite the fact that higher-mass satellites fell in more recently, as we show below.

\subsubsection{Specific angular momentum} %

Last, we present trends in specific angular momentum in the bottom row of Figure~\ref{fig:dynamics}
The median specific angular momentum, $\ell$ (bottom), decreases across time since infall from $\approx 3$ to $1 \times 10^4 \kpc\kmsi$ (bottom left panel).
Between $\MWinfall \approx 2 - 6 \Gyr$, $\ell$ is nearly constant at $\approx 2 \times 10^4 \kpc\kmsi$, which as we will show in Figure~\ref{fig:peri_dn} corresponds to satellites that completed $1 - 2$ pericentres.
The median $\ell$ is much higher for satellites that are to the left of the $t_{\rm ff}$ band, consistent with the higher velocities in these satellites.
Similar to the top row, we define the reference host halo angular momentum, $L_{\rm 200m} = V_{\rm 200m} \times R_{\rm 200m}$, and we show the range of values across hosts in the gray horizontal band.
As expected, essentially all satellites have $\ell < L_{\rm 200m}$, and most have $\ell$ significantly lower.

Generally, $\ell$ and its scatter increase with increasing $d$ as expected given that $\ell = \upsilon_{\rm tan} d$, where $\upsilon_{\rm tan}$ is the satellite's tangential velocity.
Because the MW-mass halo grew over time, satellites fell into their MW-mass halo on larger orbits at later times, which explains the increasing $\ell$ with more recent infall times.

The median $\ell$ is nearly constant with stellar mass at $\Mstar<10^{8.25}\Msun$, but as with velocity, it decreases for higher-mass satellites.
Although we find little dependence of median $d$ with $\Mstar$ (not shown), satellites with $\Mstar > 10^8 \Msun$ today exist at $\lesssim 250 \kpc$, whereas there are lower mass satellites out to $\approx 800 \kpc$.
Thus, because dynamical friction likely drove higher-mass satellites to smaller velocities and distances, they have smaller orbital lifetimes and smaller $\ell$ today.
Across the full mass range, the mean of the median is $1.6\times10^4\kpc\kmsi$, and the mean range of the 68th percentile is $0.9-2.5\times10^4\kpc\kmsi$

\begin{figure*}
\centering
\begin{tabular}{c}
\includegraphics[width=0.95\linewidth]{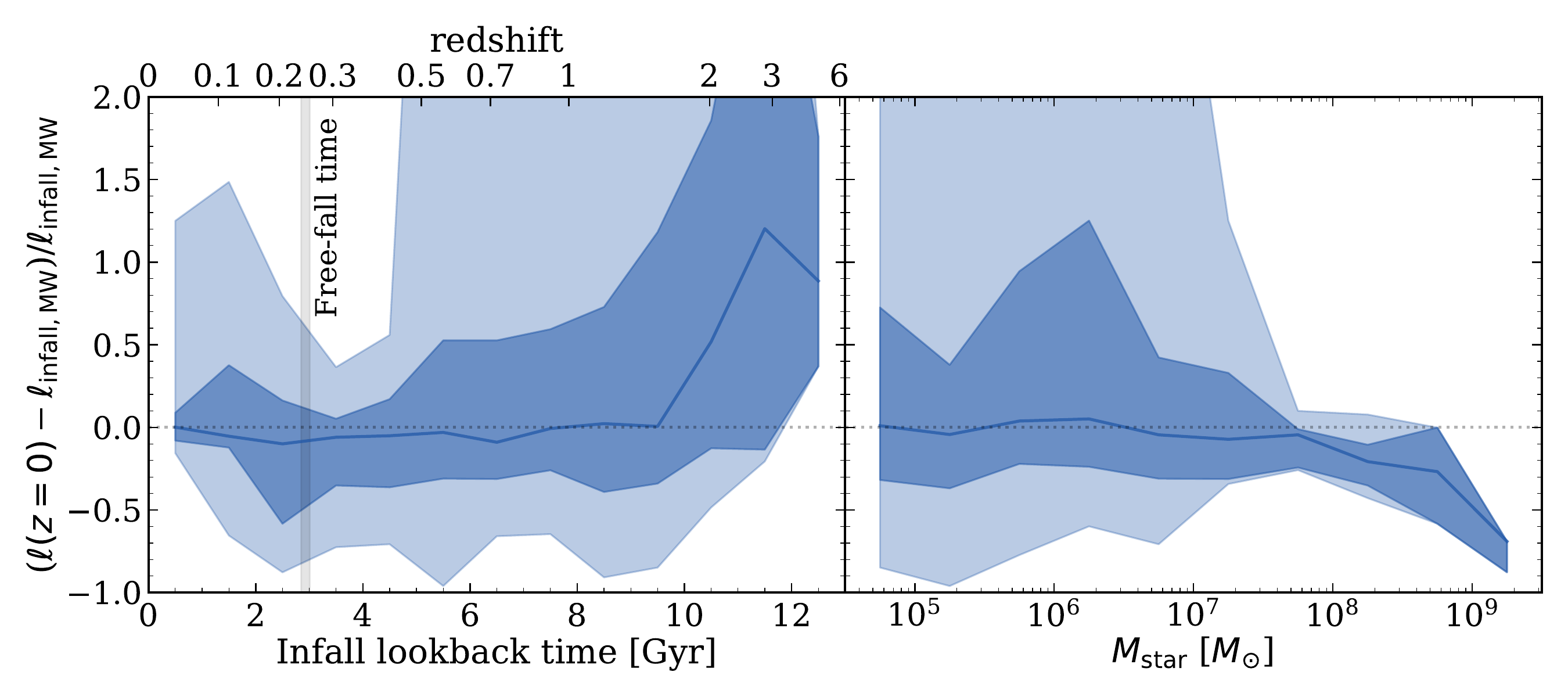}
\end{tabular}
\vspace{-3 mm}
\caption{
The fractional difference in specific angular momentum since infall into the MW-mass halo, $(\ell_{\rm now} - \ell_{\rm infall,MW}) / \ell_{\rm infall,MW}$.
Lines show the median, and the dark and light shaded regions represent the 68th percentile and full distributions, respectively.
Similar to Figure~\ref{fig:ell_evo}, the vertical gray band shows the free-fall time at the host virial radius.
\textbf{Left}: The median fractional difference in $\ell$ is within $\approx 10$ per cent for satellites that fell in $\MWinfall \lesssim 10 \Gyr$ ago, and the typical scatter about this median is $\approx 40$ percent.
However, earlier-infalling satellites \textit{increased} $\ell$, on average, up to twice their value at infall.
\textbf{Right}: Lower-mass satellites show small typical changes in $\ell$ since infall, though again with a typical scatter of $\approx 30 - 50$ per cent.
At $\Mstar \gtrsim 10^8$, satellites decreased in $\ell$, likely from stronger dynamical friction.
\textit{Thus, specific angular momentum of satellites is reasonably conserved on average, but with large ($\approx 40$ percent) scatter, for satellites with $\Mstar \lesssim 10^8$ that fell in $\lesssim 10 \Gyr$ ago.
Higher-mass satellites, or those that fell in earlier, can experience significant changes in $\ell$.}
}
\label{fig:ell_evo}
\end{figure*}

To investigate how $\ell$ changed over time, we measure the fractional difference in $\ell$ from first infall into the MW-mass halo to present-day, that is, $(\ell(z=0) - \ell_{\rm infall,MW}) / \ell_{\rm infall,MW}$.
Figure~\ref{fig:ell_evo} shows this evolution versus $\MWinfall$ (left) and $\Mstar$ (right).
The median $\ell$ did not significantly change, \textit{on average}, for satellites that fell in $\lesssim 10 \Gyr$ ago.
However, earlier-infalling satellites systematically \textit{increased} their angular momenta, by up to a factor of 2 on average, so angular momentum is not conserved for long periods in a dynamic cosmological halo environment, in part because the halo potential evolves on the same timescale as the orbit.
We also stress that the typical $1-\sigma$ width of the distribution is $\approx 40$ percent, so this represents the typical amount of dynamical scattering that a satellite experienced.
Thus, the ensemble satellite population that fell in $\lesssim 10 \Gyr$ ago did not change in $\ell$ much, but any \textit{individual} satellite's angular momentum changed by up to $\approx 40$ percent, on average.

Figure~\ref{fig:ell_evo} (right) shows that the median fractional difference in $\ell$ is minimal for satellites with $\Mstar \lesssim 10^8\Msun$, but again, the $1-\sigma$ width of the distribution is $30 - 50$ percent.
Higher-mass satellites experienced a stronger reduction in $\ell$, by roughly 70 per cent, likely from the stronger dynamical friction they experienced.

Figure~\ref{fig:dynamics} thus highlights the strong dependencies of total velocity, $E$, and $\ell$ with the lookback time of infall and present-day galactocentric distance, and a lack of dependence on mass, except at $\Mstar\gtrsim10^8\Msun$.
Our results in the middle column especially highlight the large distribution of these properties when selecting satellites at a given distance, and thus, we caution in solely interpreting the median values.
Figure~\ref{fig:ell_evo} highlights how earlier infalling and higher-mass satellites experienced larger changes in their specific angular momenta over time, as we explore below.

\subsection{Orbital histories} %
\label{sec:orbits}             %

In all but the most ideal conditions, the orbits of satellites change over time.
Mechanisms such as the time-dependent (growing) MW-mass host potential, triaxiality of the host halo, dynamical friction, and satellite-satellite interactions can perturb satellite orbits.
Because higher-mass satellites experience stronger dynamical friction, and because the MW-mass galaxy and host halo grew over time, a common expectation is that satellite orbits shrink over time, such that their most recent pericentre generally should be their smallest \citep[for example][]{Weinberg86, Taylor01, Amorisco17}.

We now explore these expectations and examine trends for the infall times and pericentres in the orbital histories of satellites at $z = 0$.
We investigate ensemble trends to help characterize and compare satellite populations in galaxies such as the MW, M31, and those in the SAGA survey.
One should interpret these results for ensembles only, and not necessarily for individual satellites and their orbits, which we will explore further in future work.

\begin{figure}
\centering
\begin{tabular}{c}
\includegraphics[width=0.95\linewidth]{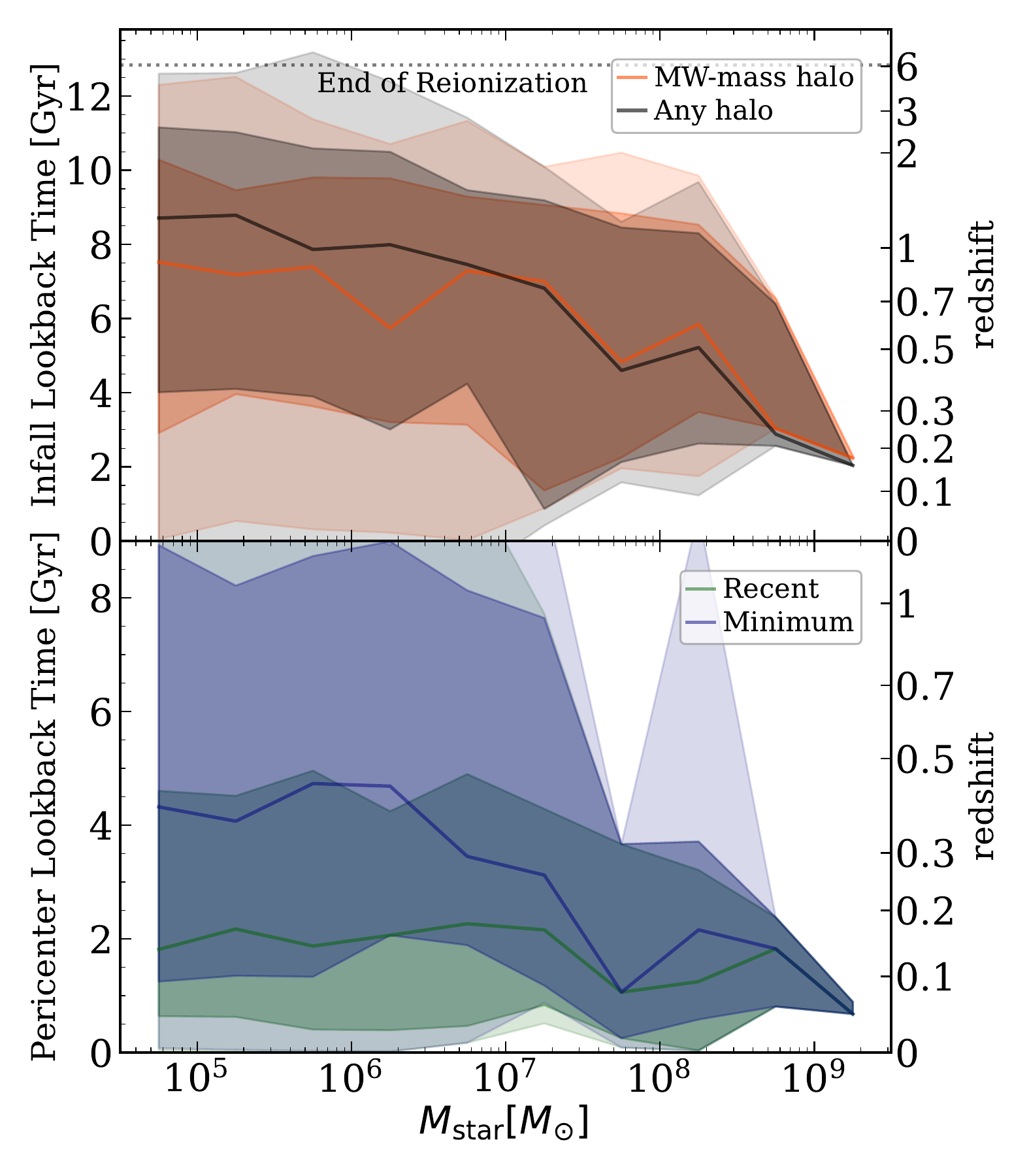}
\end{tabular}
\vspace{-3 mm}
\caption{
Lookback times to key events in the orbital histories of satellite galaxies at $z = 0$ versus their stellar mass, $\Mstar$.
Lines show the median, and the dark and light shaded regions show the 68th percentile and full distribution across the 13 hosts.
We show a dotted horizontal black line at $z = 6$ to represent the end of reionization.
\textbf{Top}: Lookback time of infall into the MW-mass halo, $\MWinfall$ (orange), and into any (more massive) halo, $\anyinfall$ (black).
For both infall metrics, higher-mass satellites fell in more recently, with $\anyinfall$ ranging from $\approx 9 \Gyr$ ago for lower-mass satellites to $\approx 2 \Gyr$ ago for higher-mass satellites, and $\MWinfall$ ranging from $\approx 7.5$ to $2 \Gyr$ ago.
Similarly, the 68th percentile and full distribution for both infall metrics span a similar time range, with the earliest infall times for any surviving satellite reaching nearly $\anyinfall \approx 13.2 \Gyr$ and $\MWinfall \approx 12.5 \Gyr$ ago.
Satellites with $\Mstar \lesssim 3 \times 10^{7} \Msun$ fell into a more massive halo typically $1 - 2 \Gyr$ before falling into the MW-mass halo, thus experiencing `group pre-processing'.
\textbf{Bottom}: Lookback time to the most recent pericentre, $\tperirec$ (green), and to the minimum pericentre, $\tperimin$ (purple), versus $\Mstar$, for satellites with at least 1 pericentre.
The median in $\tperimin$ decreases from $4 - 5 \Gyr$ ago for lower-mass satellites to $1 - 2 \Gyr$ ago for higher-mass satellites.
Conversely, the median in $\tperirec$ is roughly constant at $1-2 \Gyr$ ago across the entire mass range.
At most masses, the minimum pericentre occurred $1 - 2 \Gyr$ \textit{before} the most recent pericentre.
}
\label{fig:times_mstar}
\end{figure}

Figure~\ref{fig:times_mstar} (top) shows the lookback times when satellites fell into the MW-mass halo, $\MWinfall$ (orange) or into \textit{any} more massive halo, $\anyinfall$ (black), as a function of satellite stellar mass, $\Mstar$.
By `any more massive halo', we specifically mean any halo that is more massive than a given satellite at the same time, which could either be a MW-mass host galaxy’s halo or the halo of a massive central galaxy.
We also show a horizontal dotted line to represent when the epoch of reionization ended \citep[for example][]{Robertson21}.

Galaxies form hierarchically, so early infalling galaxies, either into the MW-mass halo or any more massive halo, were less massive.
Additionally, higher-mass satellites experienced stronger dynamical friction, which caused their orbits to lose $\ell$ and merge with the MW-mass host more quickly.
Infall lookback time into any more massive halo decreases from $\anyinfall \approx 9 \Gyr$ ago for satellites with $\Mstar \approx 10^{4.75} \Msun$ to $\anyinfall \approx 2 \Gyr$ ago for $\Mstar \approx 10^{9.25}\Msun$.
At $\Mstar \lesssim 10^7 \Msun$, satellites typically fell into another (more massive) halo before they fell into the MW-mass halo.
Above this mass, the limited range in mass between the satellite and the MW-mass halo does not leave room for an intermediate-mass host halo.
Both the 68th percentile and full distribution for each infall metric span similar ranges, in particular, the 68th percentile ranges from $4 - 11 \Gyr$ for satellites at $\Mstar < 10^7 \Msun$ and $1 - 9.5 \Gyr$ at higher mass.

The earliest infalling satellites that survive to $z = 0$ fell into a more massive halo around $\anyinfall \approx 13.2 \Gyr$ ago ($z \approx 8.6$) and into the MW-mass halo at $\MWinfall \approx 12.5 \Gyr$ ago ($z \approx 4.7$).
Furthermore, $\approx 2 / 3$ of the satellites that fell into their MW-mass halo within the first $3 \Gyr$ belong to the LG-like paired hosts, presumably because of the earlier assembly of halos in LG-like environments \citep[see][]{Santistevan20}.
Similar to the analysis of dark matter-only simulations in \citet{Wetzel15}, no surviving satellites at $z = 0$ were within their MW-mass halo before the end of the epoch of reionization at $z \gtrsim 6$ \citep{Robertson21}.
In their analysis, less than 4 per cent of satellites were a member of a more massive halo as early as $\approx 13.2 \Gyr$ ago ($z \approx 8.6$), near to when reionization was $\approx 50$ per cent complete \citep[see for example][]{FaucherGiguere20}, compared to $<1$ per cent of the satellites in our sample.
Thus, \textit{reionization was finished before surviving satellites first became satellites of a more massive halo.}
For a detailed study on satellite quenching after infall in the FIRE simulations, see \citet{Samuel22}.

Roughly $37$ per cent of satellites below $\Mstar<10^7\Msun$ were pre-processed before falling into their MW-mass hosts.
Using the ELVIS DMO simulation suite of 48 MW-mass hosts, \citet{Wetzel15} found that for pre-processed subhalos, the typical halo mass of the more massive halo they fell in was within $M_{\rm host halo} = 10^{10-12}\Msun$, with a median mass of $10^{11}\Msun$.
From Figure~\ref{fig:mstar_v_mhalo}, this corresponds to a median stellar mass of $10^{7-9}\Msun$.
\citet{Wetzel15} also determined that $\sim30$ groups contributed all pre-processed satellites, for a typical MW-mass system, with each group contributing between $2-5$ satellites.
In our sample, satellites that were pre-processed typically fell into halos with $M_{\rm halo} = 10^{9.3-11}\Msun$ before falling into the MW-mass host halo, with a median mass at infall of $M_{\rm halo} = 10^{10.2}\Msun$.
The stellar masses of galaxies hosted within these more massive halos ranged from $\Mstar\sim10^{6.4-8.9}\Msun$, with a median stellar mass of $\Mstar=10^{7.9}\Msun$.
Thus, our results are consistent with \citet{Wetzel15}, however, we see a slightly less massive central halo mass.

Figure~\ref{fig:times_mstar} (bottom) shows the lookback times to pericentres about the MW-mass host, for satellites that have experienced at least one pericentre.
Some satellites have orbited their MW-mass host multiple times, so we show the lookback time to two pericentres: the most recent pericentre, $\tperirec$ (green), and the minimum-distance pericentre, $\tperimin$ (purple).
The mass dependence of $\tperirec$ is weak.
However, lower-mass satellites experienced earlier $\tperimin$ than higher-mass satellites, again because lower-mass satellites fell in earlier, when the MW-mass halo was smaller, so their overall orbit and pericentre distance was smaller than for higher-mass satellites that fell in at later times on larger orbits.
Given that dynamical friction tends to shrink the orbits of satellites over time, particularly those with $\Mstar \gtrsim 10^8 \Msun$ or $\Mhp \gtrsim 3 \times 10^{10} \Msun$, one might assume that, for satellites that have experienced multiple pericentres, the minimum pericentre should be equal to the most recent.
However, for satellites with $\Mstar \lesssim 3 \times 10^7 \Msun$, the minimum pericentre occurred $1 - 3 \Gyr$ earlier than the most recent pericentre, and the 68th percentile in $\tperimin$ spans a much larger range in lookback time, $0.25 - 9 \Gyr$, than $\tperirec$, $0 - 5 \Gyr$.
At $\Mstar \gtrsim 3 \times 10^7 \Msun$, where the typical number of pericentres experienced is only about 1 (see below), and where dynamical friction is more efficient, the two are more comparable.
Thus, the naive expectation that low-mass satellite orbits remain relatively unchanged because they do not experience strong dynamical friction cannot explain the trends in Figure~\ref{fig:times_mstar}.
Furthermore, although the medians between $\tperimin$ and $\tperirec$ are comparable, the 68th percentile range (and full distribution) shows that differences between these pericentre metrics exist for satellites with $\Mstar \gtrsim 3 \times 10^7 \Msun$ also, implying that even the orbits of massive satellites can increase over time.
The differences between the most recent and minimum pericentres implies that some mechanism increases the pericentre distances of (especially lower-mass) satellites over time, as we explore below.

\begin{figure}
\centering
\begin{tabular}{c}
\includegraphics[width=0.95\linewidth]{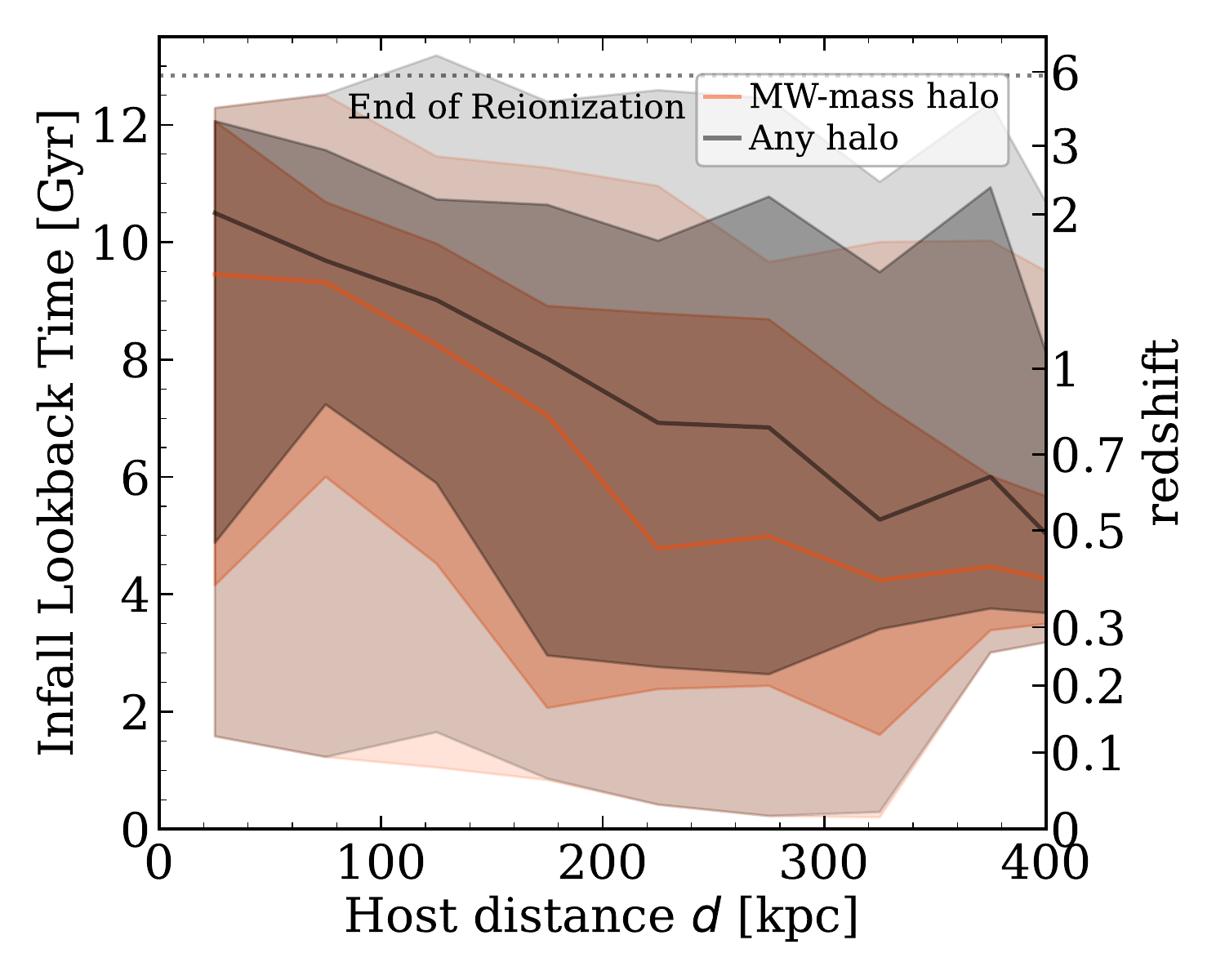}
\end{tabular}
\vspace{-3 mm}
\caption{
Similar to Figure~\ref{fig:times_mstar} (top), but showing lookback time of infall into the MW-mass halo (orange) and into any (more massive) halo (black) versus present-day galactocentric distance, $d$.
Lines show the median, and the dark and light shaded regions show the 68th percentile and full distribution across the 13 hosts.
Satellites currently closer to the host typically fell in earlier, though with significant scatter.
The lookback times of infall for the closest satellites are $10 - 10.75\Gyr$ ago, and these times decrease to $4.5 - 5.5 \Gyr$ ago for the farthest.
At any given $d$ today, satellites fell into any more massive halo $0.5 - 2 \Gyr$ before they fell into the MW-mass halo.
While $d$ correlates with infall time, we emphasize the large scatter at a given $d$, which limits the ability of using $d$ to infer infall time for any individual satellite.
}
\label{fig:infall_dz0}
\end{figure}

Figure~\ref{fig:infall_dz0} shows trends in $\MWinfall$ (orange) and $\anyinfall$ (black) with  present-day distance from the host, $d$.
Satellites currently at closer distances typically fell into another halo earlier than more distant satellites.
The closest low-mass satellites fell into any more massive halo roughly $\anyinfall \approx 10.5 \Gyr$ ago, and this median time since infall decreases to $4.8 \Gyr$ ago for satellites at $d \gtrsim R_{\rm 200m}$.
Comparing this to the time since infall into their MW-mass halo, the median $\MWinfall$ is roughly $0.5 - 2 \Gyr$ later across all distances.
The range of both the 68th percentile and full distribution generally span similar lookback times, with $\anyinfall$ offset to earlier times.
The trend of more recent time since infall at larger $d$, is because at earlier times, the MW-mass halos were smaller and satellites fell in on smaller orbits.

Again, one should not interpret our results for individual satellites but rather for populations of satellites.
Focusing on the full distribution, which extends across the full range in $d$, galaxies that fell into their MW-mass hosts between $\MWinfall \approx 3 - 9 \Gyr$ ago currently orbit at all distances between $25 - 400 \kpc$.
Thus, although the median shows a clear trend with $d$, the range in the 68th percentile is $\gtrsim 2 \Gyr$, which limits the ability to use the present-day distance of a given satellite to infer its time since infall.

Because of the mass dependence in satellite infall times in Figure~\ref{fig:times_mstar}, we checked for possible mass dependence of infall time with $d$ by splitting the sample into lower-mass ($\Mstar < 10^7 \Msun$) and higher-mass satellites.
The difference between $\anyinfall$ and $\MWinfall$ exists at all satellite distances in the low-mass sample, and because the stellar mass function is steep, we saw nearly identical results to Figure~\ref{fig:infall_dz0}.
However, the higher-mass sample showed little to no difference between the two metrics, with times since infall ranging from $3.5 - 7 \Gyr$ ago, because there were not many other more massive halos for higher-mass satellites to fall into before the MW halo.

\begin{figure*}
\centering
\begin{tabular}{c}
\includegraphics[width=0.95\linewidth]{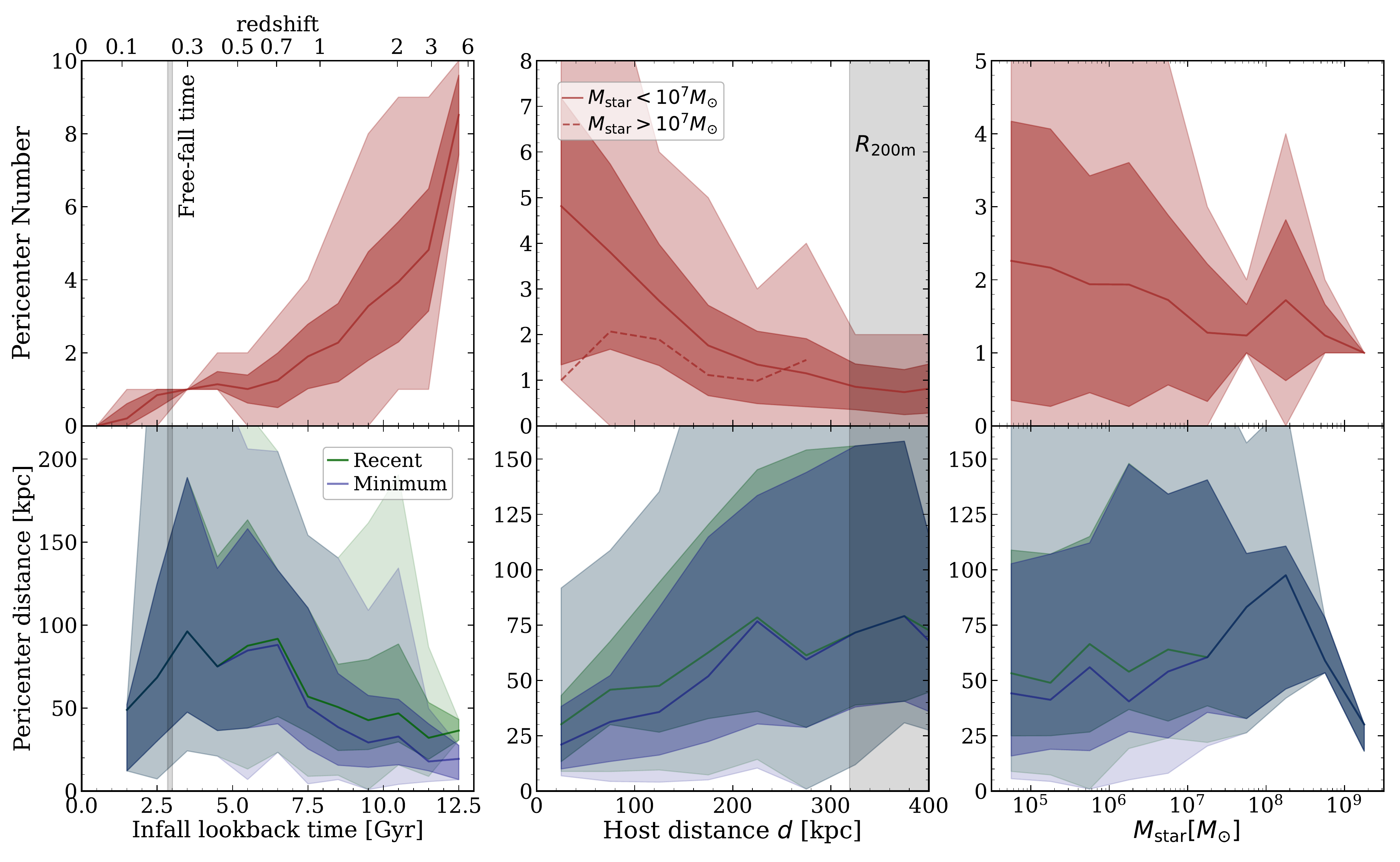}
\end{tabular}
\vspace{-3 mm}
\caption{
For satellite galaxies at $z = 0$, the number of pericentric passages about their MW-mass host, $\Nperi$ (top), and pericentre distance (bottom) versus their lookback time of infall into the MW-mass halo, $\MWinfall$ (left), present-day galactocentric distance, $d$ (middle), and their stellar mass, $\Mstar$ (right).
We present the median (mean) trends in pericentre distance (number) in the curves, and show the 68th percentile and full distribution via the dark and light shaded regions, respectively.
We include all satellites in the top row, but in the bottom row we include only satellites that have completed at least one pericentre.
Similar to Figure~\ref{fig:dynamics}, the vertical gray shaded bands show the free-fall time, $t_{\rm ff}$, at $R_{\rm 200m}$ (left) and the MW-mass halo radii, $R_{\rm 200m}$ (middle).
\textbf{Top row}: The mean $\Nperi$ increases with $\MWinfall$, where the most recently infalling satellites have not had enough time to complete a full orbit (left).
This increase of $\Nperi$ for earlier infall is because satellites had more time to orbit, and because those that fell in earlier orbit at smaller distances.
Lower-mass satellites experienced more pericentres (right), especially those currently at smaller distances (middle), given their earlier infall and the lack of strong dynamical friction acting on them.
\textbf{Bottom row}: Satellites that fell into their MW-mass host earlier experienced smaller minimum pericentres, $\dperimin$ (purple), and most recent pericentres, $\dperirec$ (green; left).
Because lower-mass satellites and satellites that are more centrally located fell in earlier, they orbit with smaller $\dperimin$ and $\dperirec$ (middle and right).
However, higher-mass satellites also feel stronger dynamical friction that causes them to merge into the host on shorter timescales.
The results highlight how satellite orbits changed between the most recent and minimum pericentres, which often do not occur at the same distance, especially for satellites with $\Mstar \lesssim 10^7 \Msun$.
}
\label{fig:peri_dn}
\end{figure*}

Figure~\ref{fig:peri_dn} shows trends in the number of pericentric passages, $\Nperi$, about the MW-mass host (top row) and various pericentric distance metrics (bottom row), versus the time since infall into MW-mass halo, $\MWinfall$ (left), present-day distance from the MW-mass host, $d$ (middle), and satellite stellar mass, $\Mstar$ (right).
Again, we include vertical gray shaded regions that represent the free-fall time at $R_{\rm 200m}$, $t_{\rm ff}$ (left column), and MW-mass halo $R_{\rm 200m}$ (middle), as reference values.
When presenting trends in $\Nperi$, we include all satellites, including those that have not yet experienced a pericentric passage, but for trends in pericentre distance we only include satellites that have experienced at least one pericentre.
Satellites that have not yet reached first pericentre comprise $\approx 7$ per cent of all satellites.

The top left panel shows the expected trend of more pericentres for earlier $\MWinfall$.
The mean $\Nperi$ is 0 for recently infalling satellites, and it rises to $\Nperi \approx 1$ and is flat across $\MWinfall \approx 2.5 - 5.5 \Gyr$ ago, because this time interval is comparable to an orbital timescale.
$\Nperi$ then rises rapidly with $\MWinfall$, reaching nearly 9 for the earliest-infalling satellites.
We compared these trends for lower-mass versus higher-mass satellites and find no significant differences.

Figure~\ref{fig:peri_dn} (top middle) shows the dependence of $\Nperi$ on $d$.
Because we find significant differences in $\Nperi$ with $\Mstar$ (top right), we split the sample into $\Mstar < 10^7 \Msun$ (solid) and $\Mstar > 10^7 \Msun$ (dashed).
We choose this mass selection given that the lower-mass satellites experience a mean $\Nperi \geq 2$, while the higher-mass satellites have a mean of $\Nperi \approx 1$.
Lower-mass satellites generally experienced more pericentres than higher-mass satellites at a given $d$, and the mean number decreases with $d$ for lower-mass satellites from $\Nperi \approx 5$ to 1 for those near $R_{\rm 200m}$ (gray shaded region).
Conversely, we do not find dependence on $d$ for higher-mass satellites, with a mean value of $\Nperi \approx 1-2$ at all $d$, likely because of their more recent infall and the increased importance of dynamical friction on them.

Finally, $\Nperi$ declines weakly with $\Mstar$, with a mean $\Nperi \approx 2.5$ at $\Mstar = 10^{4.75} \Msun$ to $\Nperi \approx 1$ at $\Mstar > 10^9 \Msun$.
Lower-mass satellites experienced more pericentres, because they fell in earlier (see Figure~\ref{fig:times_mstar} top), and also because higher-mass satellites took longer to form and felt stronger dynamical friction that caused them to merge into their MW-mass host on shorter timescales.
Over the full sample, the largest number of pericentres experienced is $\Nperi = 4$ at $\Mstar > 10^7 \Msun$, and $\Nperi = 10$ for lower-mass satellites.

The bottom row shows trends for both pericentre metrics: the pericentre with the minimum distance, $\dperimin$ (purple), and the most recent pericentre, $\dperirec$ (green).
In the idealized scenario we outline at the beginning of this subsection, an orbiting satellite's pericentre will remain unchanged or shrink over time because of dynamical friction.
However, the panels above show that this often is not true; early infalling satellites can have larger subsequent pericentres.

The bottom left panel shows the median $\dperimin$ (purple) and $\dperirec$ (green).
Both $\dperimin$ and $\dperirec$ are nearly identical for satellites that fell in $\MWinfall < 6 \Gyr$ ago, with median values ranging from $60 - 100 \kpc$.
For earlier-infalling satellites, both distance trends slowly decrease from $\approx 100 \kpc$ to $20 - 35 \kpc$, where the most recent pericentre was roughly $5 - 20 \kpc$ larger than the minimum.
Thus, for sufficiently early-infalling satellites which spent longer amounts of time in the evolving MW-mass host halo, the orbits grew slightly over time, which is not expected given the assumption of unchanged orbits or shrinking orbits due to dynamical friction.

We also investigated how the first pericentre a satellite experienced depends on infall time and find qualitatively similar results to the other two pericentre metrics.
Earlier-infalling satellites had smaller first pericentres than later-infalling satellites, and the first pericentres were smaller than the most recent ones.
The first pericentres were also the minimum a satellite ever experienced in a majority ($\approx 72$ per cent) of satellites with $\Nperi \geq 2$.
The only noticeable differences between the first pericentres and $\dperimin$ occurred for galaxies that fell in $\gtrsim 6 \Gyr$ ago.

The bottom middle panel shows pericentre distance trends versus current $d$.
As expected, both pericentre metrics increase with $d$, from $20 - 30 \kpc$ for satellites that are currently closer, to $70 - 80 \kpc$ for satellites near $R_{\rm 200m}$ (gray shaded region).
Satellites within $d \lesssim 225 \kpc$ often had recent pericentres that are larger than $\dperimin$ by nearly $10 - 20 \kpc$, so the orbits of these satellites grew.
The pericentre metrics in both lower-mass and higher-mass satellites increase with $d$, but because lower-mass satellites fell in earlier than higher-mass satellites and completed more orbits, they largely drive the differences between $\dperimin$ and $\dperirec$ (and $\dperimin$ and $d_{\rm peri,first}$) in the bottom left panel.
We again highlight that the full distributions in both $\dperimin$ and $\dperirec$ span a wide range at a given $d$, so even though the median trend increases with $d$, one should not directly apply our results to an individual satellite.

Finally, the bottom right panel shows that lower-mass satellites typically had smaller recent and minimum pericentres.
However, at $\Mstar \gtrsim 10^{8.25} \Msun$, the median pericentre distances decrease, likely driven by the onset of efficient dynamical friction.
Lower-mass satellites have smaller pericentre distances because they fell-in earlier when the MW-mass halo was smaller and less massive.
The typical recent/minimum pericentre distance is $40 - 60 \kpc$ for satellites with $\Mstar \lesssim 10^7 \Msun$, $60 - 100 \kpc$ for satellites with $\Mstar \approx 10^{7-8.25} \Msun$, and $\lesssim 100 \kpc$ for higher-mass satellites.

Because the mass of the host can determine the orbits of the satellites, we investigated potential differences between satellites in higher-mass and lower-mass host halos.
At pericentre, satellites are deep in the potential near the galaxy, therefore, the stellar mass of the central galaxy could also correlate with our pericentre metrics.
Specifically, we divided the sample in two by selecting the 6 MW-mass hosts with the higher $\Mstar$ and 7 hosts with lower $\Mstar$ (see Table~\ref{tab:hosts}) and examined their pericentre distances versus $\MWinfall$ and satellite $\Mstar$.
We find no differences between the two samples versus $\Mstar$.
Versus $\MWinfall$, the satellites in higher-mass host halos had slightly larger, although minimal, $\dperimin$ and $\dperirec$.

The results in Figures~\ref{fig:times_mstar}-\ref{fig:peri_dn} suggest a different evolution than expected for some satellites.
Lower-mass satellites fell into their MW-mass hosts earlier, when the halo was smaller and less massive, so they complete more orbits than higher-mass satellites in this evolving potential and orbit at smaller distances.
Interestingly, the orbits of these lower-mass satellites can increase over time, presumably through the evolving global potential or interactions with other galaxies, which opposes the common expectation of shrinking orbits.
However, given the 68th percentile ranges and the full distribution of the pericentre properties, differences between the most recent and minimum pericentres exist at \textit{all} satellite masses, and not solely at low mass.

\subsection{Satellites with growing pericentres} %
\label{sec:torqued}                              %

As we showed in Figures~\ref{fig:times_mstar}-\ref{fig:peri_dn}, the most recent pericentre that a satellite experienced is often not the minimum in terms of distance.
We now investigate these cases in more detail and refer to satellites with $\dperimin < \dperirec$ as having `growing pericentres'.

\begin{figure}
\centering
\begin{tabular}{c}
\includegraphics[width=0.95\linewidth]{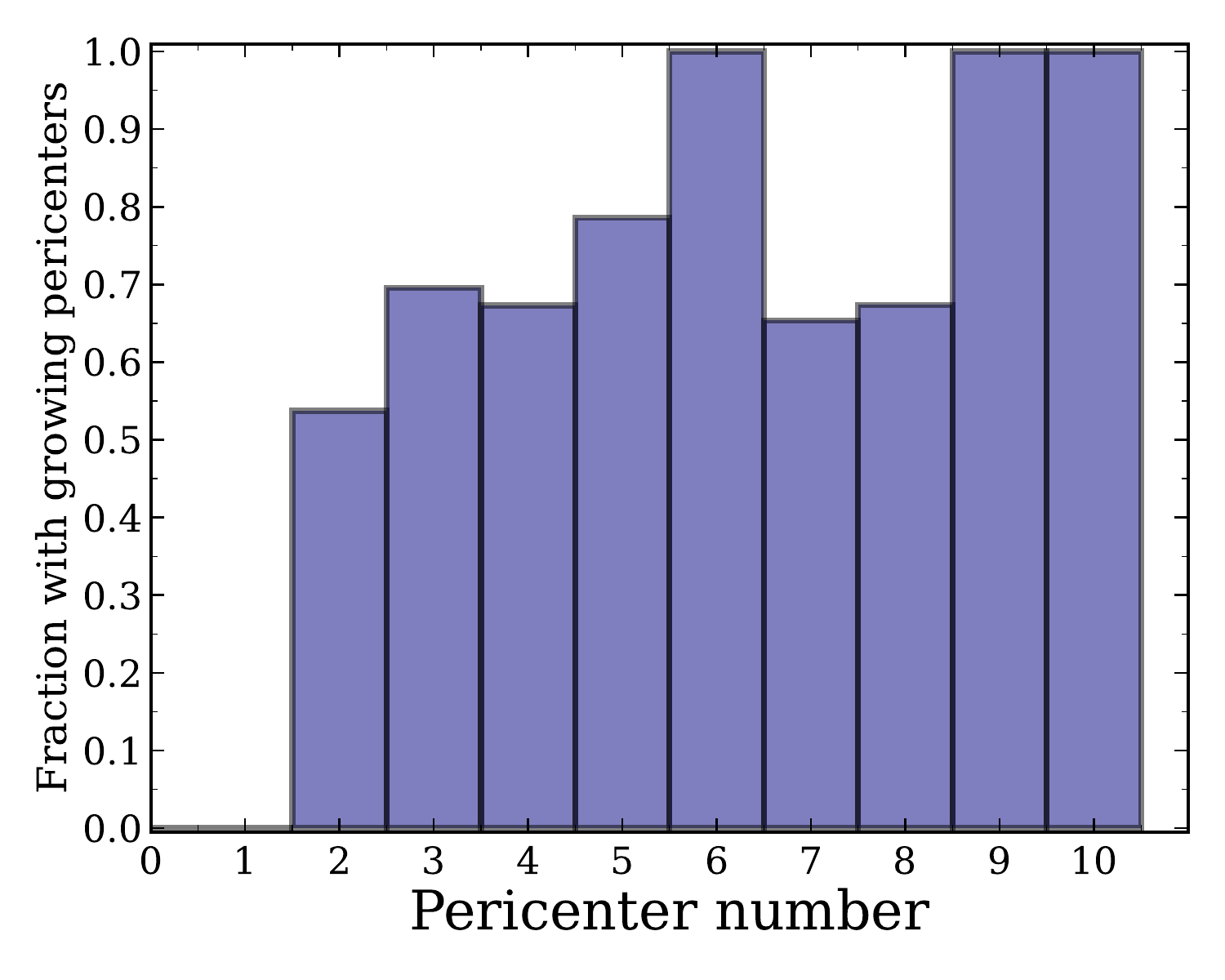}
\end{tabular}
\vspace{-4 mm}
\caption{
Fraction of satellite galaxies with growing pericentres, relative to all satellites that experienced the given number of pericentres, $\Nperi$.
By definition, these satellites must have $\Nperi \geq 2$, which is why the fraction is zero for $\Nperi = 0$ and 1.
\textit{Satellites with growing pericentres represent the majority of all satellites with $\Nperi \geq 2$.}
}
\label{fig:fraction}
\end{figure}

Satellites with growing pericentres make up 31 per cent of all satellites (ranging from $23 - 46$ per cent for a given host).
Moreover, growing pericentres comprise the \textit{majority} (67 per cent across all hosts, ranging from $50 - 86$ per cent for a given host) of all satellites with $\Nperi \geq 2$.
In other words, \textit{for satellites with two or more pericentres, typically their most recent pericentre was not their closest encounter with their MW-mass host galaxy}.
Figure~\ref{fig:fraction} highlights this, showing the fraction of satellites with growing pericentres versus pericentre number.
For satellites with $\Nperi \geq 2$, the growing pericentre population represents $> 50$ per cent of the total sample at any $\Nperi$, and in some cases, they represent the \textit{entire} population at a given $\Nperi$.
This fraction broadly increases with $\Nperi$, at least up to $\Nperi = 6$, where it represents all satellites, though the fraction fluctuates for $\Nperi$ above that.

Although we cannot directly check whether or not this is a temporary occurrence, we compared the most recent pericentre distance to the maximum pericentre a satellite experienced.
For satellites that experienced more than 3 pericentres, we found that roughly 30 per cent of them experienced their maximum pericentre sometime between the minimum and most recent.
Of these satellites, the fractional difference between their most recent pericentre distance, and their maximum, i.e. $(d_{\rm peri,recent}-d_{\rm peri,max})/d_{\rm peri,max}$, was between 1-54 per cent, with a median value of 17 per cent, and a 68th percentile range of 8-38 per cent.
Thus, because this scenario happens in the minority of satellites, and because the median fractional difference is small, we argue that it is not merely a temporary occurrence.
Furthermore, from the top right panel of Figure~\ref{fig:peri_dn}, satellites with more than 3 pericentres are generally lower-mass satellites, which will not strongly feel the effects of dynamical friction.

We confirmed that this population of satellites with growing pericentres is not sensitive to the choice for the center of the MW-mass host in computing satellite distances.
Specifically, we examined these trends using the center of the host dark-matter halo (instead of the center of the stars in the host galaxy, as is our default).
This results in only 7 additional satellites whose minimum and most recent pericentres differ by more than 5 per cent, which represents only $\approx 4$ per cent of all satellites with $\Nperi \geq 2$.

\begin{figure*}
  \begin{minipage}{.33\linewidth}
    \centering
    \includegraphics[width=.95\linewidth]{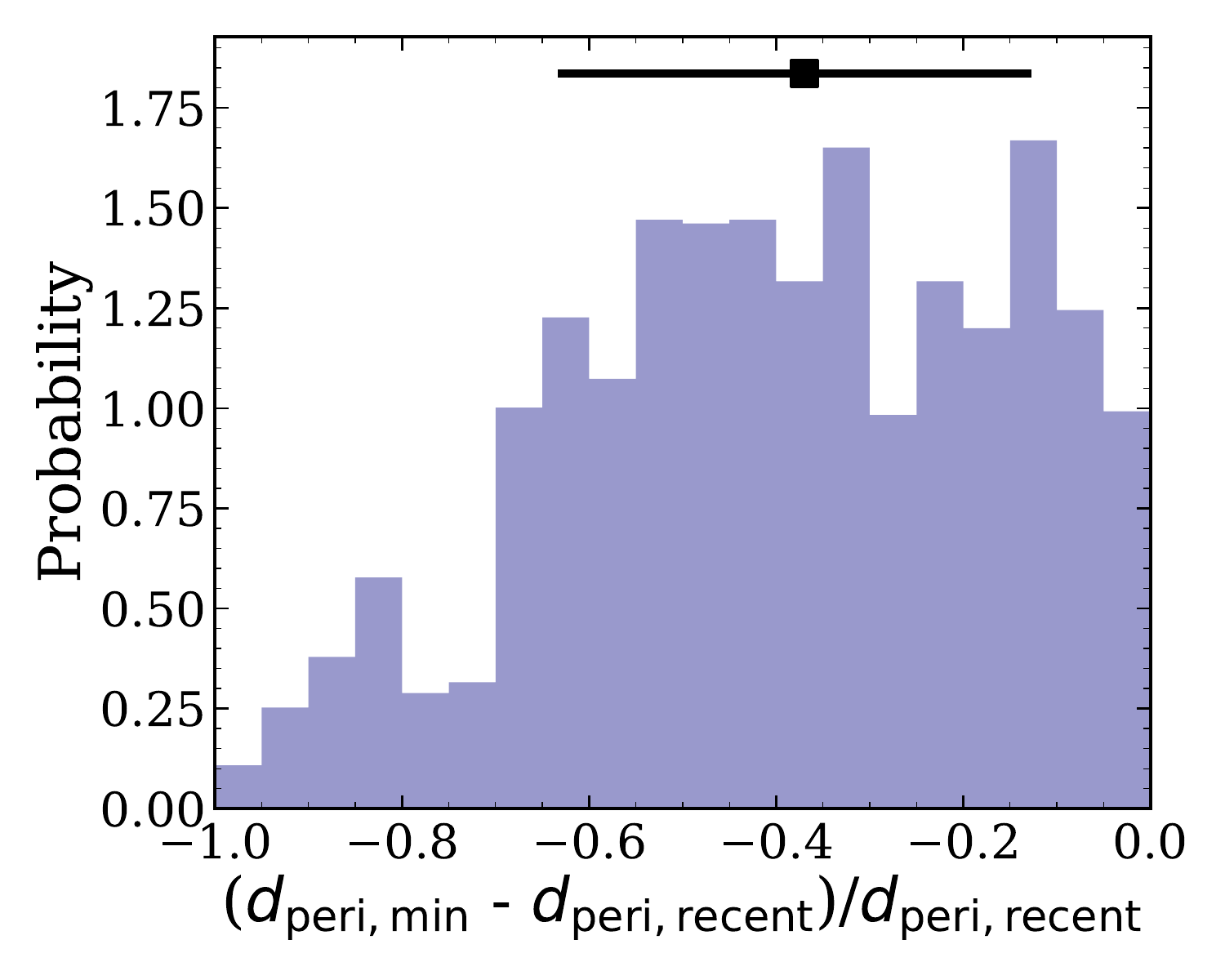}\par
  \end{minipage}%
  \begin{minipage}{.33\linewidth}
    \centering
    \includegraphics[width=.95\linewidth]{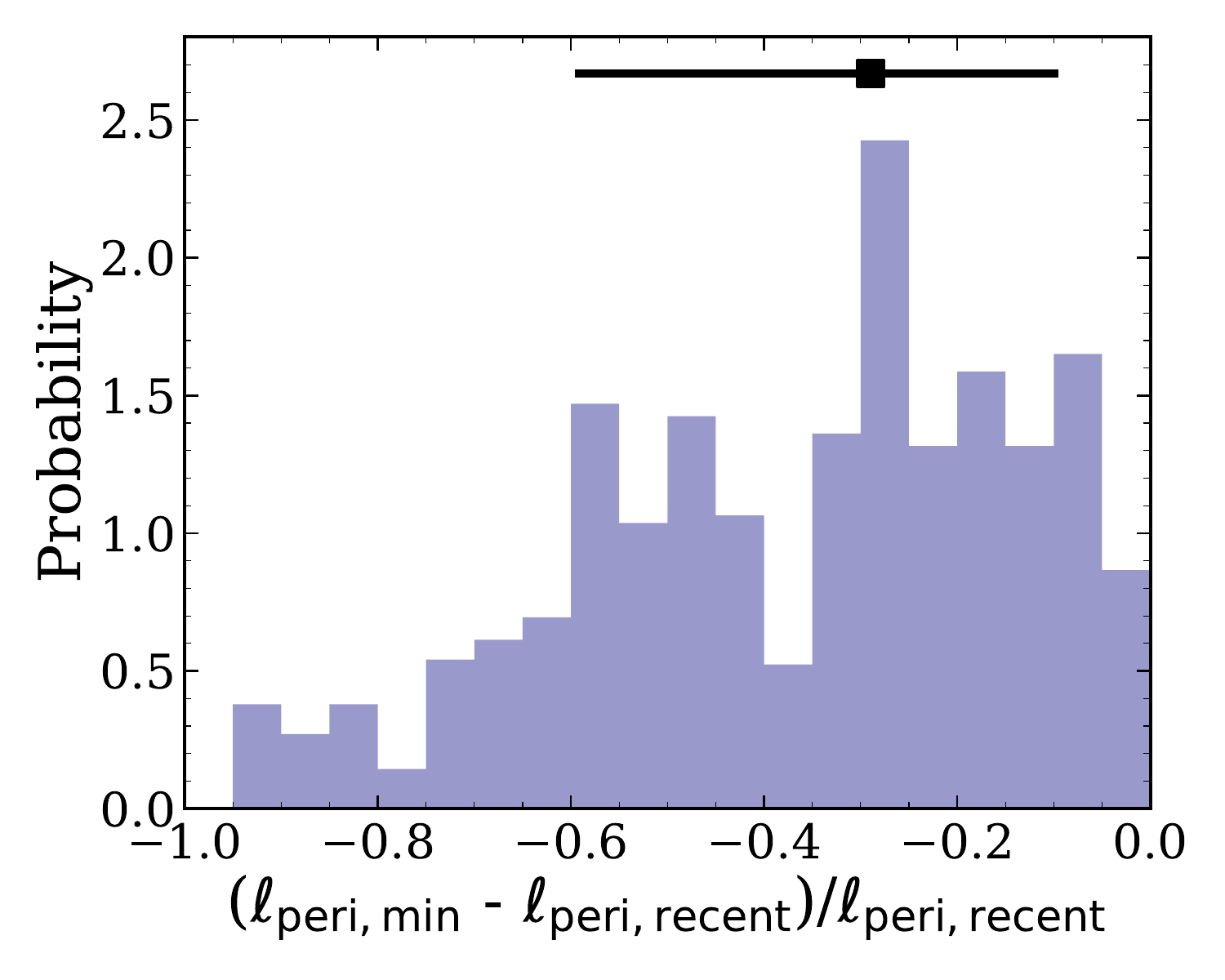}\par
  \end{minipage}%
  \begin{minipage}{.33\linewidth}
    \centering
    \includegraphics[width=.95\linewidth]{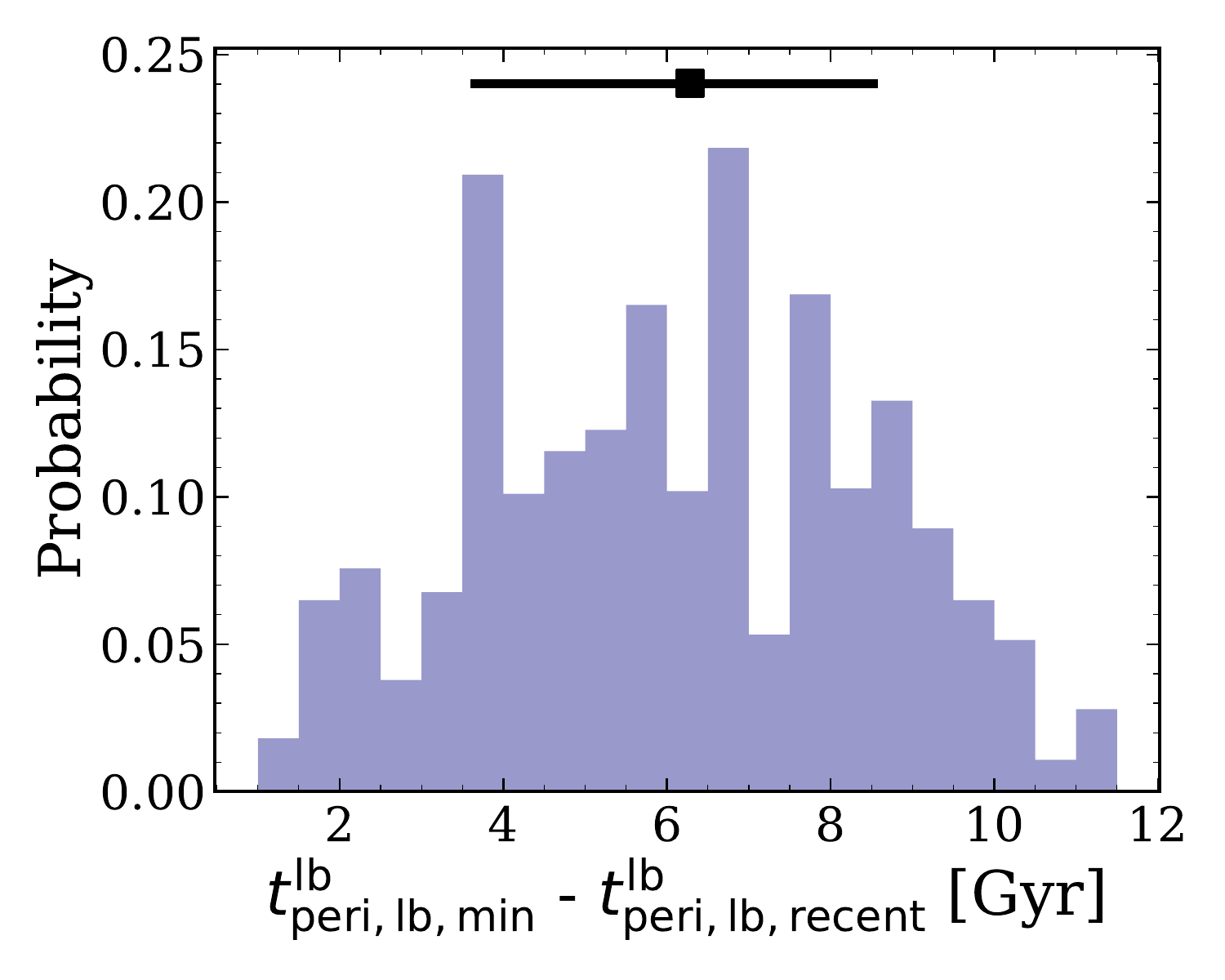}\par
  \end{minipage}%
\vspace{-2 mm}
\caption{
Probability distributions of changes in orbital properties for satellite galaxies with growing pericentres.
This population comprises the majority (67 per cent) of all satellites that experienced multiple pericentres, or 31 per cent of satellites overall.
Each panel shows the median value and 68th percentile via the black square point with error bars.
\textbf{Left}: The fractional difference between the minimum and recent pericentre distances, $(\dperimin - \dperirec) / \dperirec$.
The median is $-37$ per cent.
\textbf{Middle}: The fractional difference between the specific angular momenta at the minimum and recent pericentres, $(\ell_{\rm peri,min} - \ell_{\rm peri,recent}) / \ell_{\rm peri, recent}$.
The median is $-29$ per cent.
\textbf{Right}: The difference between the lookback times to the minimum and recent pericentre, $\tperimin - \tperirec$.
The median is $\approx 6.3 \Gyr$, or $1 - 2$ orbits, given the free-fall time.
\textit{All metrics show significant differences between the most recent and the minimum pericentre for a given satellite.}
}
\label{fig:torqued_hist}
\end{figure*}

To quantify further the significance of this population, Figure~\ref{fig:torqued_hist} shows the probability distributions of the difference between key properties at the minimum and most recent pericentres for all satellites with growing pericentres.
The left panel shows the fractional difference between the two pericentre distances, $(\dperimin - \dperirec) / \dperirec$.
As the black point shows, the median fractional difference is $-37$ per cent, with a 68th percentile range of $-15$ to $-65$ per cent.

Figure~\ref{fig:torqued_hist} (middle) shows the fractional difference in specific angular momentum at the minimum and most recent pericentres.
Nearly all satellites with growing pericentres ($> 95$ per cent) experienced an increase in $\ell$ between the two pericentres; we do not show in Figure~\ref{fig:torqued_hist} the small percent with increased $\ell$.
The median fractional difference in $\ell$ is $-29$ per cent, with a range in the 68th percentile of $-10$ to $-60$ per cent.

\begin{figure*}
\centering\includegraphics[width=.99\linewidth]{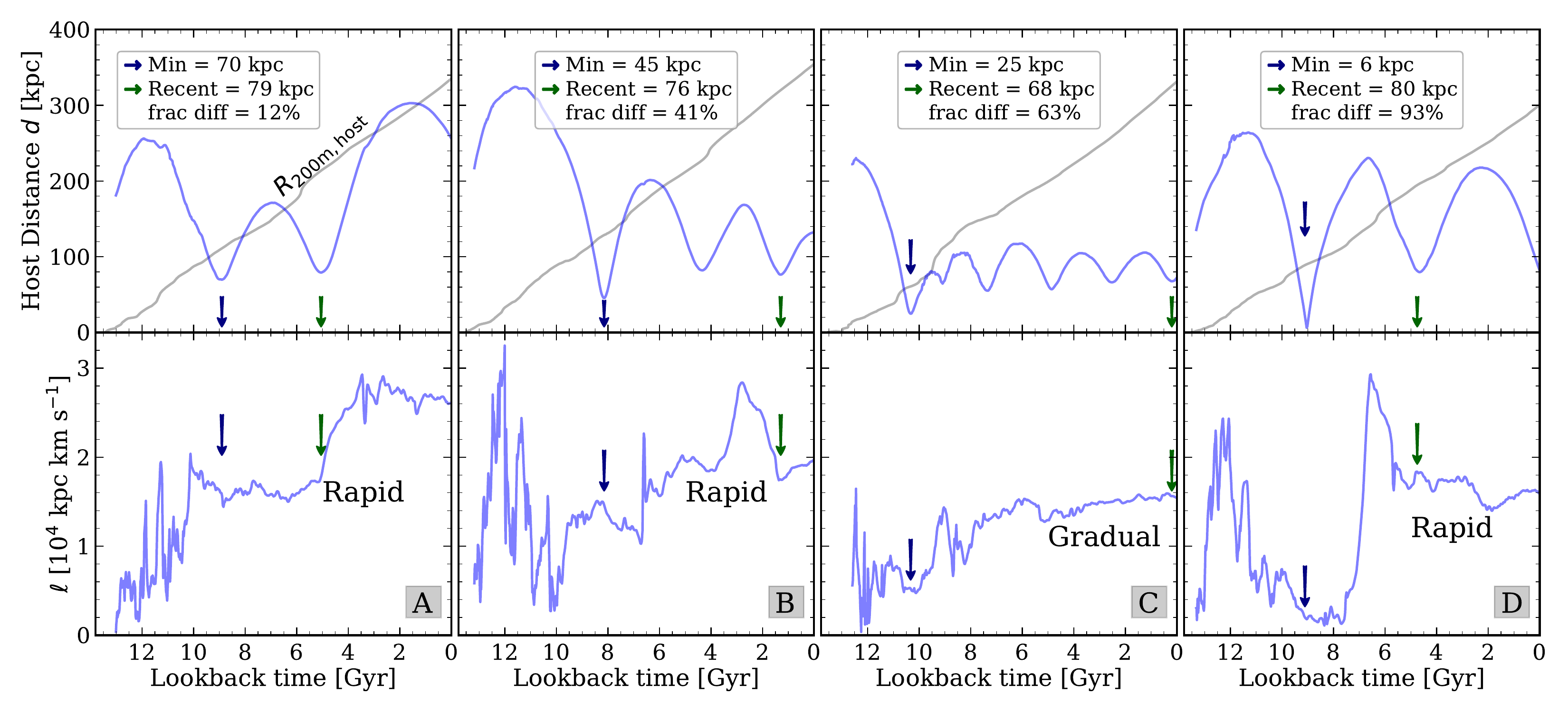}\par
\vspace{-3 mm}
\caption{
Orbital distance, $d$ (top), and specific angular momentum, $\ell$ (bottom), for 4 representative satellites with growing pericentres, labeled A-D.
For each satellite, we list its minimum and recent pericentre distances, along with the fractional difference between the two, and the purple and green arrows show when these events occurred.
The top row also shows the growth of the MW-mass $R_{\rm 200m}$ (grey).
The majority of satellites with growing pericentres (69 per cent) experienced $\dperimin$ within $1 \Gyr$ after infall, in 72 per cent of satellites with growing pericentres the first pericentre was the minimum, and nearly 71 per cent of this satellite population orbited beyond the MW-mass host $R_{\rm 200m}$ after their first pericentre.
The evolution of $\ell$ shows that the increase in pericentre distance can happen rapidly or gradually, but that it often does not happen near pericentre.
\textit{Approximately 30 per cent of cases show a sharp increase in $\ell$ during the first apocentre after infall, which suggests that these satellites may be interacting with other halos or non-axisymmetric features in the density field.}
}
\label{fig:torqued_orbits}
\end{figure*}

Finally, Figure~\ref{fig:torqued_hist} (right) shows the difference between the lookback times of the minimum and most recent pericentres.
These satellites have a wide range of time differences, with a median of $\approx 6.3 \Gyr$ and 68th percentile range of $3.5 - 8.5 \Gyr$.
These are slightly longer than the typical orbital periods of these satellites, $2 - 5 \Gyr$, as Figure~\ref{fig:torqued_orbits} shows, the minimum and most recent pericentres do not always occur successively.

To provide more context, Figure~\ref{fig:torqued_orbits} shows the orbits (host distance versus time) for four representative satellites with growing pericentres (top row), along with their specific angular momentum (bottom row), labeled A-D from left to right.
We chose these four particular satellites at random to span the entire possible range of fractional pericentre distances.
The legends show the values of the minimum and most recent pericentres, along with the fractional differences between them; these four satellites range from $12 - 93$ per cent.
The arrows indicate when these pericentres occurred.
For reference, we also show the MW-mass halo's $R_{\rm 200m}(t)$ (grey).

All four satellites experienced $\dperimin$ immediately after first infall, and the first pericentre was the minimum in 72 per cent of all satellites with growing pericentres.
Furthermore, as Figure~\ref{fig:torqued_orbits} suggests, $71$ per cent of satellites with growing pericentres experienced a splashback phase of orbiting beyond $R_{\rm 200m,host}$ after their first pericentre.
For comparison, among the population with $N_{\rm peri} > 1$ but $\dperimin = \dperirec$, only $57$ per cent experienced a splashback phase.
So this suggest that orbiting beyond $R_{\rm 200m,host}$ is associated with a growing pericentre, at least in some cases.

As Figure~\ref{fig:torqued_orbits} (bottom row) suggests, nearly all satellites whose pericentres grew also increased their specific angular momentum.
By visually inspecting the histories of the full population, we find that this occurs in two broad ways: (1) steady, gradual increase in $\ell$ over time, which accounts for 45 per cent of all growing pericentres, and (2) rapid growth in $\ell$ near a pericentre or apocentre, which account for 53 per cent of the satellites.
The remaining 2 per cent of satellites are the rare cases in which the pericentres increased from minimum to the most recent, but the angular momentum decreased.
The fractional change in $\ell$ for these satellites is generally small, $\lesssim6$ per cent.
However, some of these satellites show clear signs of interactions with other galaxies, and fell in early ($\gtrsim 8.5 \Gyr$ ago).
For satellites in category (1), a time-dependent and/or triaxial host halo potential likely plays an important role, especially given that satellites with growing pericentres typically fell in early, when the shape of the host halo potential was changing more rapidly \citep[for example][Baptista et al, in prep]{Santistevan20, Gurvich22}.
Satellite C in Figure~\ref{fig:torqued_orbits} shows a relatively gradual increase in $\ell$ over time.
We defer a more detailed investigation to future work.

For the growing pericentres in category (2), $4/5$ of the satellites experienced a rapid increase in $\ell$ near either a single apocentre, or some combination of them, with the first apocentre being the most common.
This is especially apparent in satellites B and D in Figure~\ref{fig:torqued_orbits}.
The other satellites showed rapid increases in $\ell$ involving a pericentre that was not the minimum pericentre, much like in satellite A.

Because the fraction of splashback orbits is higher for satellites with growing pericentres compared to the remaining population with multiple pericentres, this suggests that perturbations at $d \gtrsim R_{\rm 200m,host}$ may play a key role in causing this population.
This behavior is apparent in satellites B and D of Figure~\ref{fig:torqued_orbits}, where large spikes in $\ell$ occur near apocentres, some of which are beyond $R_{\rm 200m,host}$.
These rapid increases in $\ell$ are caused by rapid increases in the tangential velocities, which typically were of order $\delta \upsilon \approx 30 \kmsi$.

Other satellite mergers with the MW-mass host also can significantly alter the global potential, resulting in orbit perturbations.
We investigated correlations of both $\tperimin$ and $\tperirec$ with the lookback times of mergers, with stellar mass ratios of $\gtrsim 1:100$, and did not find a clear correlation between these times.
We also investigated correlations of these pericentre metrics with various metrics of host formation times including: the lookback times of when the host galaxy formed 90 per cent of its stellar mass \citep[see][for a table of values]{Gandhi22}, the lookback times of when the host formed 10 per cent of its halo mass, and the lookback time of when the host galaxy's growth transitioned from being dominated by mergers to in-situ formation \citep{Santistevan20}.
We find no significant correlations with these formation metrics.

\begin{figure*}
\centering
\begin{tabular}{c c}
\includegraphics[width=0.475\linewidth]{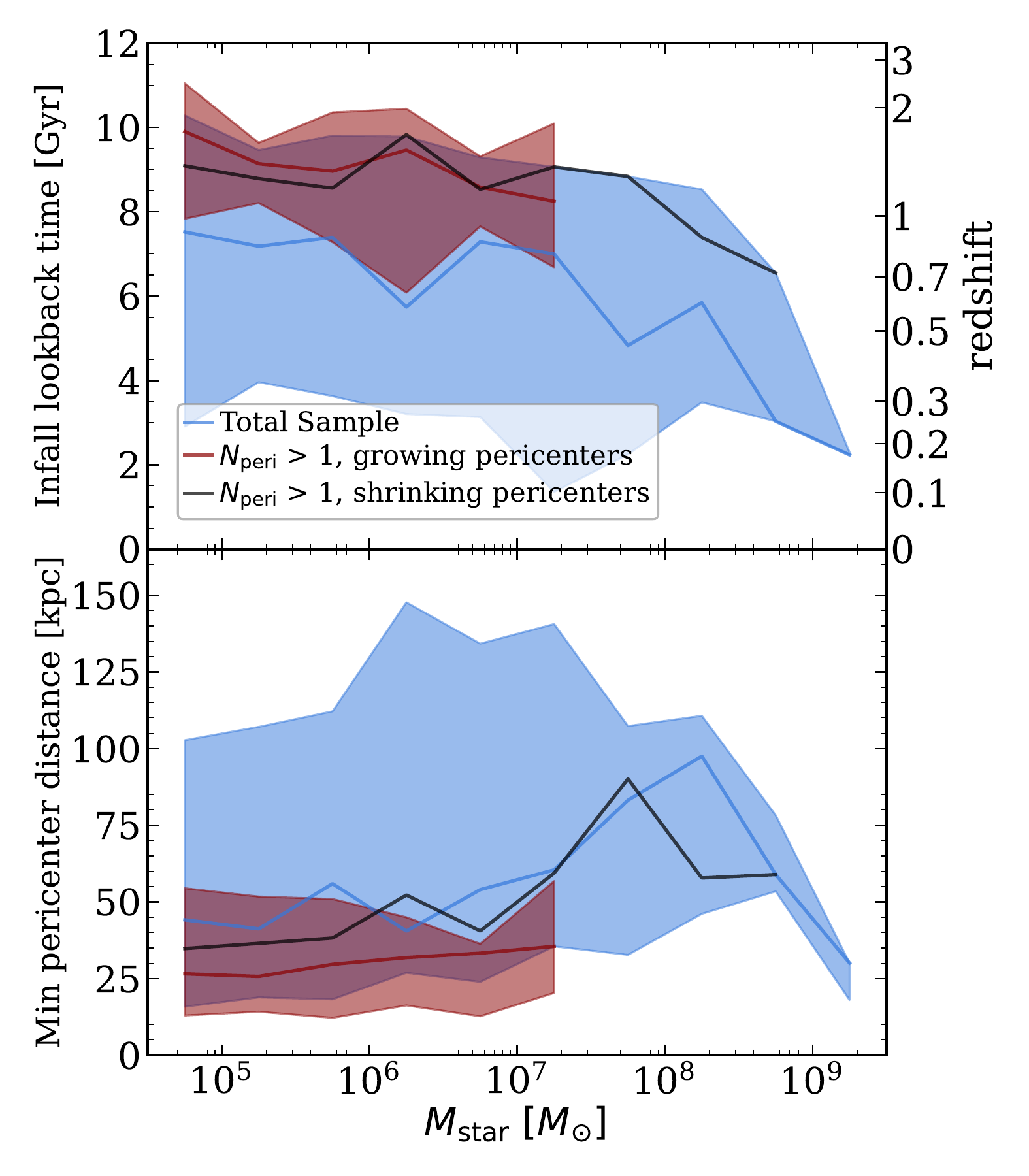}&
\includegraphics[width=0.475\linewidth]{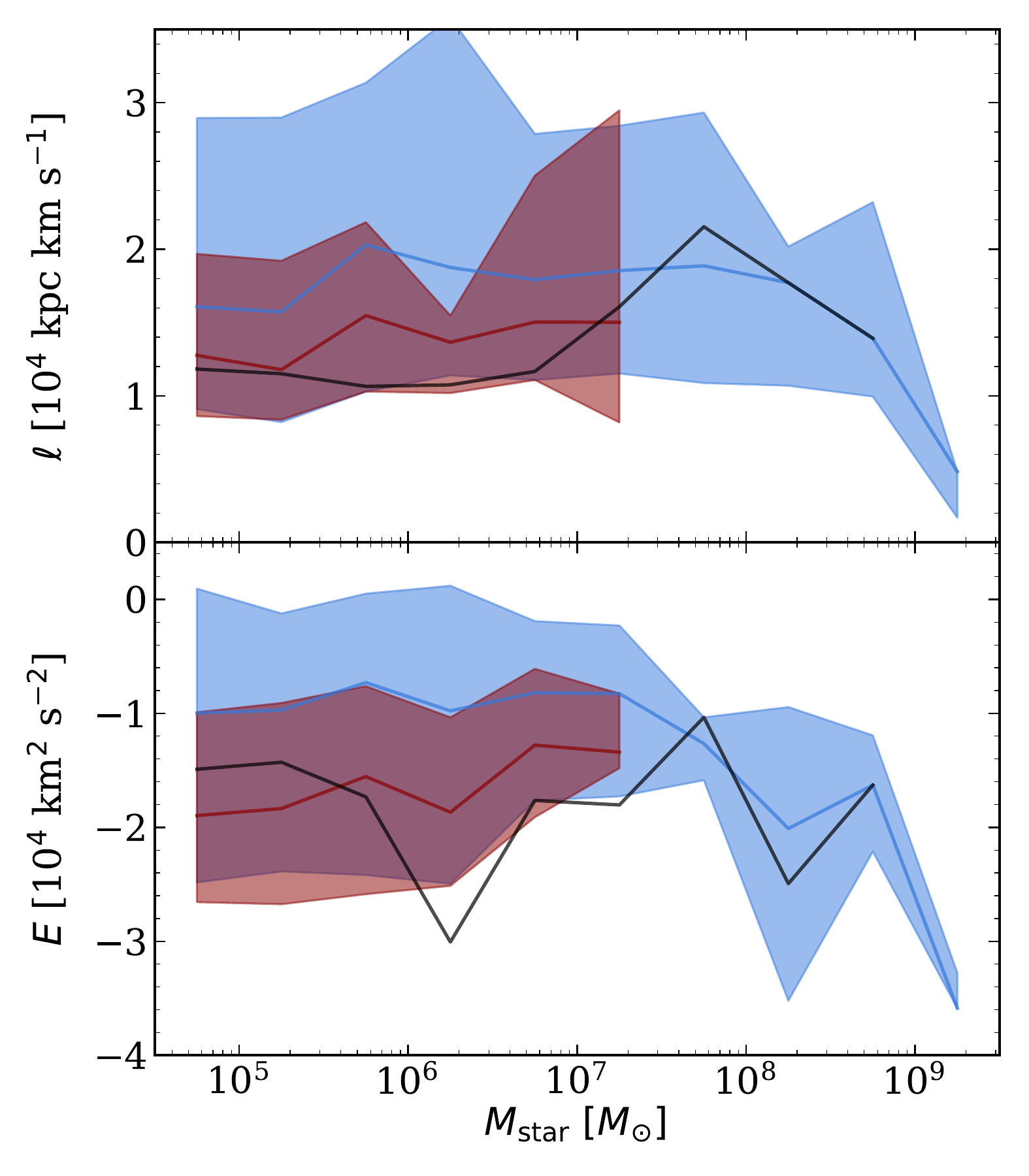}
\end{tabular}
\vspace{-5 mm}
\caption{
Orbital properties of satellite galaxies at $z = 0$, including the total sample (blue), those whose minimum and most recent pericentre distances are not the same (`$\Nperi > 1$, growing pericentres' in red), and those that experienced multiple pericentres but excluding the `growing pericentre' population (`$\Nperi > 1$, shrinking pericentres' in black).
Solid lines show median values, and shaded region shows the 68th percentile range for the growing pericentre and total population.
\textbf{Top left}: Satellites that experienced multiple pericentres necessarily fell in earlier than the total sample, but among those with $\Nperi > 1$, the growing pericentre population is not systematically biased in infall time.
\textbf{Bottom left}: Satellites with growing pericentres experienced $10 - 25 \kpc$ smaller $\dperimin$, while the remaining satellites with shrinking pericentres experienced $\dperimin$ more similar to the total sample.
\textbf{Top right}: Because satellites with $\Nperi > 1$ fell into the MW-mass halo earlier than the total sample (when the host $R_{\rm 200m}$ was smaller), they have smaller $\ell$ today, and among these, satellites with growing pericentres have slightly higher $\ell$ today given their selection.
More massive satellites with $\Nperi > 1$ have larger $\ell$, closer to the total sample, because they fell in more recently.
\textbf{Bottom right}: Because satellites with $\Nperi > 1$ fell into the MW-mass halo earlier than the total sample and orbit at smaller distances, they are more bound today.
Satellites with growing pericentres have orbital energies today comparable to the other satellites with $\Nperi > 1$, though slightly higher at intermediate masses.
In summary, satellites with growing pericentres fell in at comparable times to other satellites with $\Nperi>1$, however, the satellites with growing pericentres orbit with smaller pericentre distances, and with slightly larger $\ell$, which makes them unique.
}
\label{fig:torqued_mstar}
\end{figure*}

To investigate whether satellites with growing pericentres have biased orbits, both throughout their history and today, Figure~\ref{fig:torqued_mstar} shows several orbital properties versus stellar mass, for satellites with growing pericentres, all satellites, and all satellites with $N_{\rm peri} > 1$ but with shrinking pericentres, that is, with $\dperimin = \dperirec$.
The top left panel shows the lookback time of infall into the MW-mass halo.
Compared to the total sample, as expected, both sub-samples with $\Nperi > 1$ fell in typically $\gtrsim 1 - 2 \Gyr$ earlier.
However, among the population with $\Nperi > 1$, we find no significant differences between those with growing versus shrinking pericentres, so infall time does not correlate with having a growing pericentre.

Figure~\ref{fig:torqued_mstar} (bottom left) shows the minimum pericentre distances, $\dperimin$.
Although all three samples show similar behaviour to Figure~\ref{fig:peri_dn}, with increasing pericentre distance with increasing $\Mstar$, the growing pericentre population is biased to the smallest $\dperimin$.
The shrinking pericentre sub-sample is generally consistent with the total sample, with typical values spanning $\dperimin \approx 35 - 90 \kpc$, while $\dperimin$ for growing pericentre satellites is $10 - 25 \kpc$ smaller, ranging from $\dperimin \approx 25 - 35 \kpc$.
Thus, satellites with growing pericentres orbited closer to the host galaxy.
Again, $\approx 30$ per cent of satellites with growing pericentres experienced rapid increases in $\ell$ during their first apocentre or slightly after $\dperimin$.
Likely, other important factors contribute to the larger differences between the pericentre metrics, such as the evolving MW-mass host potential, gravitational interactions with other satellite galaxies, and mergers, Figures~\ref{fig:torqued_orbits} hints at, as we plan to explore in future work.

Figure~\ref{fig:torqued_mstar} (top right) shows the specific angular momentum at $z = 0$.
Because the growing and shrinking pericentre sub-samples fell into their MW-mass halo earlier, we expect them to have smaller $\ell$.
However, the growing pericentres having modestly higher $\ell$ at $z = 0$ at most masses, again reflecting that they have scattered to larger $\ell$ by today.

Figure~\ref{fig:torqued_mstar} (bottom right) shows the specific orbital total energy, $E$.
Consistent with their earlier infall, both sub-samples with $\Nperi > 1$ are on more bound orbits today than the total population, at least at $\Mstar < 10^{7.25} \Msun$.
Any systematic differences between the growing and shrinking pericentres are modest, given the scatter, so we conclude that there are no clear differences in specific orbital total energy at $z = 0$.

A satellite galaxy can undergo significant mass stripping when it orbits throughout the MW-mass host halo, especially when it is deepest in the host's potential at pericentre, and this drastic loss in the satellite's subhalo mass subsequently can affect its orbit.
Thus, to better understand the origin of satellites with growing pericentres, including the timescales over which the orbits changed and potential dynamical perturbations near $\dperimin$, we compared the specific angular momentum and DM subhalo mass $200 \Myr$ before and after the minimum pericentre.
Near $\dperimin$, $\ell$ changed by a much smaller amount ($10 - 20$ per cent) than the change in $\ell$ from the minimum to most recent pericentres ($\approx 40$ per cent).
The fractional mass lost near $\dperimin$ was also minimal ($\lesssim 7$ per cent).
Thus, in general, the orbital perturbations did not occur just near $\dperimin$, as also apparent in Figure~\ref{fig:torqued_orbits}.

Finally, we investigated the other orbital properties we presented in the previous figures (total velocity, pericentre lookback times, and the recent pericentre distances) for these sub-samples but find no compelling differences.
We find no mass dependence to the fractional differences in pericentre distances or times in Figure~\ref{fig:torqued_hist}, so satellites with growing pericentres exist similarly at a range of masses.
Thus, even though higher-mass satellites experience stronger dynamical friction and have smaller orbital lifetimes, we find no mass dependence to whether or not a satellite has an orbit with growing pericentres.
We find no strong correlation of the fractional distance or time metrics with either $\dperimin$, $\dperirec$, or the lookback times that these occurred.
Unsurprisingly, the fractional difference in pericentre distance increased slightly with earlier $\MWinfall$, given that satellites that orbited for a longer amount of time, had more time to experience changes in their orbits.

In summary, of all satellites that experienced $\Nperi \geq 2$, the majority (67 per cent) experienced a growing pericentre.
The most recent pericentre distance is typically $\approx 37$ per cent higher than the minimum experienced, which occurred $\sim 6 \Gyr$ earlier.
Interestingly, about half (45 per cent) of growing pericentres experienced a gradual increase in $\ell$, presumably from a time-dependent and/or triaxial MW-mass host potential, and about half (53 per cent) experienced rapid growth in $\ell$ following either their first or minimum pericentres, during their first apocentres, or during multiple pericentre or apocentre events, which suggests a perturbation by another galaxy.
Satellites with growing pericentres are more likely to have been splashback satellites, further suggesting perturbations at large distances.
Furthermore, because we measure the orbits of these satellites relative to the MW-mass host galaxy, another effect may be perturbations to the center of mass of the host galaxy from mergers or massive satellites.
Given these complexities and likely multiple causes for the origin of satellite with growing pericentres, we defer a more detailed investigation to future work.

\section{Summary \& Discussion} %
\label{sec:sumndisc}            %

\subsection{Summary of Results} %
\label{sec:sum}                 %

We investigated the orbital dynamics and histories of 473 satellite galaxies with $\Mstar > 3 \times 10^4 \Msun$ around 13 MW-mass galaxies from the FIRE-2 suite of cosmological simulations.
Surprisingly, and in contrast to many (semi)analytical models of satellite evolution, most satellites that experienced multiple orbits experienced an increase in orbital pericentre and specific angular momentum, likely from interactions with the MW-mass host or other satellites.
This highlights that satellite orbits do not always shrink and that angular momentum is not always conserved throughout a satellite's orbital history.

In summary, the topics that we presented in the Introduction and our corresponding results are:

\renewcommand{\labelenumi}{\arabic{enumi}}
\renewcommand{\labelenumii}{\arabic{enumi}.\arabic{enumii}}
\begin{enumerate}
\item \textit{The relation of orbital properties of satellite galaxies at $z = 0$ to their orbital histories, including lookback times of infall, distances from the MW-mass host, and stellar masses.}
\begin{itemize}
    \item Satellites that fell in earlier have lower orbital energies and specific angular momenta, \textit{though with significant scatter}, because satellites that fell in earlier necessarily had to be on smaller orbits to be captured by the MW halo, and the MW-mass host potential continued to grow over time (Figure~\ref{fig:dynamics}).
    \item Satellites closer to the host generally orbit with higher velocities, smaller specific angular momenta, and have more bound orbits, \textit{though with significant scatter} (Figure~\ref{fig:dynamics}).
    Total velocity, specific angular momentum, and specific orbital energy do not correlate with $\Mstar$ except at $\Mstar \gtrsim 10^8 \Msun$, where dynamical friction is more efficient (Figure~\ref{fig:dynamics}).
    \item \textit{Specific angular momentum, $\ell$, often is not conserved, even approximately, throughout a satellite's orbital history} (Figure~\ref{fig:ell_evo}).
    In particular, earlier-infalling satellites \textit{increased} in $\ell$ since infall.
    More expectedly, higher-mass satellites decrease in $\ell$, likely because of dynamical friction.
    The range of fractional changes in $\ell$ at smaller $\Mstar$ and later infall extends $\gtrsim 50$ per cent.
    That said, the average $\ell$ across the full satellite population remains statistically unchanged since infall.
    \item Many lower-mass satellites were pre-processed before becoming a satellite of the MW-mass host.
    At $\Mstar < 10^7 \Msun$, 37 per cent fell into another more massive halo \textit{before} falling into the MW-mass halo, typically $\approx 2.7 \Gyr$ before (Figures~\ref{fig:times_mstar}, \ref{fig:infall_dz0}, \ref{fig:times_mhalo}).
    \item \textit{No surviving satellites were within the MW-mass halo during the epoch of reionization} ($z \gtrsim 6$), and less than 4 per cent were satellites of any host halo during this time, similar to \citet{Wetzel15} (Figures~\ref{fig:times_mstar} and \ref{fig:times_mhalo}).
    Surviving satellites at $z = 0$ fell into the MW-mass halo as early as $12.5 \Gyr$ ago, and into any host halo as early as $13.2 \Gyr$ ago.
    \item Satellites at a given distance today experienced a large range of infall times into the MW-mass halo.
    Thus, one cannot infer a precise infall time based solely on a satellite's present-day distance alone, and the use of total velocity, specific angular momentum, or specific orbital energy alone is similarly limited (Figures~\ref{fig:dynamics}, \ref{fig:infall_dz0}).
\end{itemize}

\textit{\item Testing a common expectation that the orbits of satellite galaxies shrink over time, that is, that a satellite's most recent pericentric distance is the minimum that it has experienced.}
\begin{itemize}
    \item Most satellites at $z = 0$ with $\Mstar \lesssim 10^7 \Msun$ experienced more than one pericentre, while more massive satellites experience only one (Figure~\ref{fig:peri_dn}), because of their later infall and dynamical friction.
    \item \textit{Contrary to the expectation that satellite orbits tend to shrink over time,
    most satellites that experienced 2 or more pericentres have grown in pericentre distance.}
    Of all satellites with $\Nperi \geq 2$, 67 percent experienced a growing pericentre.
    This represents 31 per cent of all satellites.
    \item Typically, the minimum percienter was $37$ per cent smaller than the most recent one, because the fractional specific angular momentum increased by $30$ per cent (Figure~\ref{fig:torqued_hist}).
    This minimum pericentre typically occurred $\sim 6 \Gyr$ before the most recent one (Figure~\ref{fig:torqued_hist}).
    \item Satellites with growing pericentres orbited closer to the host ($\dperimin = 24 - 35 \kpc$) than those with shrinking pericentres.
    \item Perturbations at large distances likely contribute to these changes in satellite orbits, given the high fraction (71 per cent) of growing pericentres that were once a splashback satellite.
    However, we find no single dynamical origin: 53 per cent of satellites with growing pericentres experienced a large increase in $\ell$ during one or more apocentre, while 45 per cent experienced a gradual, steady increase in $\ell$.
    This suggests that as the MW-mass host halo grows over time, this may help slowly torque the satellites to larger orbits, such that their subsequent pericentres increase.
    We leave a more detailed investigation of this to future work.
\end{itemize}
\end{enumerate}

\subsection{Inferring Infall Times from Present-day Properties} %
\label{sec:infall}                                              %
We presented various trends of present-day properties, such as total velocity, specific angular momentum, $\ell$, specific energy, $E$, and distance from the host galaxy, $d$, with the lookback time of satellite infall, $\MWinfall$.
The median trends in these present-day properties often correlate with $\MWinfall$.
However, we stress that distribution of infall times at fixed property span a large range, limiting the ability to use a property like present-day distance to infer the infall time of a single satellite.

For example, in Figure~\ref{fig:dynamics}, while the median specific energy decreases with increasing $\MWinfall$, for a satellite with a specific energy of $E = -1 \times 10^4 \kmsis$, the 68 per cent range in $\MWinfall$ is $1.5 - 10.5 \Gyr$ ago.
Similarly, although the median specific angular momentum decreases with increasing $\MWinfall$, a satellite with $\ell = 2 \times 10^4 \kpc\kmsi$ fell in $1 - 9 \Gyr$ ago.
Figure~\ref{fig:infall_dz0} shows that, for a satellite at $100 \kpc$ today, $\MWinfall \approx 5.5 - 10.5 \Gyr$, and for a satellite near the host virial radius, $d \approx 300 \kpc$, it experienced $\MWinfall \approx 2 - 8 \Gyr$ ago.

Furthermore, across Figures~\ref{fig:dynamics} and \ref{fig:infall_dz0}, at a given satellite total velocity, $\ell$, $E$, or $d$, the \textit{full} distribution of infall times spans $\approx 13 \Gyr$, nearly the age of the Universe.
Thus, while these figure show trends in the median for a population of satellite galaxies, we caution that using any of one of these present-day properties for a single satellite will not precisely determine its infall time into the MW-mass halo.
In future work, we will explore how precisely one can infer infall time using full 6D phase-space information, including knowledge about the host potential.

\subsection{Comparison to Previous Work}        %
\label{sec:sim_comp}                            %

First, we re-emphasize that these FIRE-2 simulations broadly reflect the observed population of satellites in the LG.
\citet{Wetzel16} and \citet{GarrisonKimmel19a} showed that their satellite stellar mass functions and internal velocity dispersions (dark-matter densities) broadly agree with the MW and M31.
\citet{Samuel20} showed that their radial distance distributions broadly agree with the MW and M31, and with MW-mass galaxies from the SAGA survey.
Furthermore, \citet{Samuel21} showed that, although uncommon, spatially or kinematically coherent planes of satellites exist in these simulations, similar to what is observed in the MW and M31.
These benchmarks are important for motivating our analysis of their satellite orbits and histories.

Our results agree with \citet{Wetzel15}, who examined similar trends of satellite infall against both $\Mstar$ and $d$, using the ELVIS suite of cosmological zoom-in DMO simulations, with abundance matching to assign stellar mass to subhalos across $\Mstar = 10^{3-9} \Msun$.
The mass and spatial resolution in these simulations were $1.9 \times 10^5\Msun$ and $140$ pc, respectively, $\approx 5 \times$ and $\approx 3 \times$ larger than our baryonic simulations.
They found that satellites typically first fell into any more massive halo $\approx 6.5-10 \Gyr$ ago and into the MW-mass halo between $\approx 5-7.5 \Gyr$ ago.
These times since infall are consistent with the top panel of Figure~\ref{fig:times_mstar} (and the results in Figure~\ref{fig:times_mhalo}) for lower-mass satellites, but the lookback times of infall for satellites at $\Mstar \gtrsim 10^7 \Msun$ in our results are more recent than in \citet{Wetzel15}: $\lesssim 6 \Gyr$ ago for both infall metrics.
As we show in Appendix~\ref{app:dmo}, the addition of baryonic physics, especially the additional gravitational potential of the MW-mass galaxy, causes stronger tidal mass stripping and disruption, and because the higher-mass satellites additionally feel stronger dynamical friction, we only see a few higher-mass satellites that happen to survive in our baryonic simulations.
As a function of distance from the MW-mass host, \citet{Wetzel15} also found that satellites experience first infall (into any more massive halo)  $\approx 4 - 11 \Gyr$ ago and infall into their MW-mass halo $\approx 3-9 \Gyr$ ago, consistent with our results in Figure~\ref{fig:infall_dz0}.

\citet{Rocha12} used the Via Lactea II DMO simulation and found a strong correlation between satellite orbital energy and infall time \citep[see also][]{Fillingham19, DSouza22}.
The authors suggest that satellites that are deeper in the gravitational potential at $z = 0$ often fell in earlier than satellites farther out and have more negative orbital energies.
We find qualitatively consistent values and dependencies of infall time with $d$ as in \citet{Rocha12}: satellites presently closer to the MW-mass galaxy fell in earlier and are on more bound orbits.

\citet{Bakels21} used one of the $N$-body Genesis simulations, which have mass and spatial resolution of $7.8 \times 10^6\Msun$ and $1.1 \kpc$, respectively, which is $\approx 200 \times$ and $\approx 30 \times$ larger than the resolution in our simulations, to study the infall histories of satellites.
They analyzed the orbits of (sub)halos of 2309 hosts with $M_{\rm 200c} \geq 0.67 \times 10^{12} \Msun$ and found that roughly 22 per cent of all subhalos are on first infall, much larger than the $\approx 7-8$ per cent in our sample.
Furthermore, \citet{Bakels21} found that roughly 60 per cent of the splashback population of halos have yet to reach their first apocentre, and the majority ($\approx 86$ per cent) have only reached pericentre once, indicating that this population of satellites are on long-period orbits.
Given the wide range in MW-mass $R_{\rm 200m}$, if we select satellites that are currently beyond $300 - 400 \kpc$ to represent the splashback population, we find comparable results: over 95 per cent of the satellites have only experienced one pericentre so far, and the remaining 5 per cent have only experienced two pericentres.
For the subhalos that have experienced pericentre at least once and are currently inside of the host's virial radius, \citet{Bakels21} found that about half of them have only experienced one pericentre, and $\approx 30$ per cent have not yet reached apocentre.
Our result in the bottom left panel of Figure~\ref{fig:peri_dn} is generally consistent with this, with a median pericentre number of $1-2$ for all satellites in our sample.
However, of satellites that experienced at least one pericentre, we find that $\gtrsim 2/3$ of them completed more than one, which suggests that the satellites in our simulations completed more orbits.
Finally, \citet{Bakels21} noted that roughly 95 per cent of the surviving subhalos were accreted since $z = 1.37$ ($9.1 \Gyr$ ago), where we generally see earlier infall: 95 per cent of our satellites fell into their MW-mass host since $z = 2.2$ ($10.7 \Gyr$ ago).
Thus, the satellites that survive to $z = 0$ in our simulations fell in earlier resulting in the larger fraction that have completed more orbits.
The differences in the first infall fractions, and the accretion times of satellite galaxies, between our results and \citet{Bakels21} are likely because of the differences in resolution between the FIRE-2 and Genesis simulations.
Because the Genesis simulations have DM particle masses $\approx200\times$ larger, they necessarily resolve only more massive satellites.

\citet{Fattahi20} used the cosmological baryonic zoom-in simulations of MW-mass galaxies from the Auriga project to investigate the $\MWinfall$ for surviving and destroyed low-mass galaxies, and their effect on the growth of the stellar halo.
They also found that surviving satellites fell into their MW-mass halos more recently than the destroyed satellites, similar to the results in \citet{Panithanpaisal21} and \citet{Shipp22}, who used the same 13 FIRE-2 simulations in our analysis to investigate stellar stream progenitors, their orbits, and their detectability.
\citet{DSouza21} also found similar results in their DMO satellite analysis.
The analysis by \citet{Fattahi20} shows similar results to the top panel in our Figure~\ref{fig:times_mstar}, with more massive satellites falling in more recently.
At $\Mstar = 10^6 \Msun$ and $\Mstar = 10^9 \Msun$, the authors report average infall lookback times of $\MWinfall = 7.8 \Gyr$ and $\MWinfall = 3.8 \Gyr$.
Our results are broadly consistent, though shifted to more recent times, where satellites at $\Mstar = 10^6 \Msun$ and $\Mstar = 10^9\Msun$ fell into their MW-mass halo with mean infall lookback times of $\MWinfall \approx 6.5 \Gyr$ and $\MWinfall \approx 2.9 \Gyr$.
We only have 3 satellites at $\Mstar \sim 10^9 \Msun$, smaller than the sample in \citet{Fattahi20}.
The differences between the infall times between satellites in FIRE-2 and Auriga may arise from differences in the stellar mass - halo mass relation at low-mass \citep{Auriga, Hopkins18}.
Furthermore, the way in which we both average satellite properties over the hosts may contribute to the differences in infall time, given that host galaxies with larger satellite populations will skew the results.

\citet{Wetzel15} concluded that, in the ELVIS suite of DMO simulations, no present-day satellites were within the MW-mass halo's virial radius during the epoch of reionization at $z \gtrsim 6$.
This implies that, for any satellites whose star formation quenched during that time, the MW environment was not the driving factor, so the effects of the MW halo environment and cosmic reionization are separable, in principle.
We similarly conclude that no satellites at $z = 0$ were within their MW-mass halo virial radius during reionization.
Although our resolution is still finite, the trend in Figure~\ref{fig:times_mstar} is relatively flat with mass, with no indication that it significantly increases for lower-mass satellites.
Also, as we show in Appendix~\ref{app:mhalo}, we find similar infall trends in subhalos down to $\Mhp = 10^8 \Msun$, which would host ultra-faint galaxies (Figure~\ref{fig:times_mhalo}).
Recently, \citet{Sand22} proposed that the ultra-faint galaxy Tucana B, whose nearest neighbor is $\approx 500 \kpc$ away, was likely quenched in an isolated environment from reionization.
It has an old ($\approx 13.5 \Gyr$), metal-poor ($\rm [Fe/H] \approx -2.5$), stellar population, and has no recent star formation.
Thus, because of its distance to any other massive galaxy, old stellar population, and lack of star formation, \citet{Sand22} argued that Tucana B is an excellent candidate for a galaxy quenched by reionization.
However, our results, and those in \citet{Wetzel15}, imply that no present-day satellites were within a MW-mass halo during reionization.
Thus, selecting isolated galaxies \textit{today} does not necessarily make them cleaner probes of the effects of reionization.
Rather, satellites around MW-mass galaxies today provide similarly good candidates to study these effects.

Using the ELVIS DMO simulations, \citet{Wetzel15} showed that many satellite galaxies first were pre-processed, for $0.5 - 3.5 \Gyr$, before falling into their MW-mass halo; $\approx 30$ per cent of satellites with $\Mstar = 10^{3-4} \Msun$ were members of another group during their infall into a MW-mass halo, and this fraction decreases to $\approx 10$ per cent at $\Mstar = 10^{8-9} \Msun$.
Any time before their infall into the MW-mass halo, $\approx 60$ per cent of low-mass satellites were members of another more massive group, falling to $\approx 30$ for high-mass satellites.
Over our full sample of satellites, nearly 35 per cent fell into another more massive halo before falling into the MW-mass host, consistent with \citet{Wetzel15}.
The fraction of pre-processed satellites in our results is also comparable to the DMO-based results from \citet{Li08} who reported that $\approx 1/3$ of subhalos were pre-processed, though they selected subhalos down to $M_{\rm halo} \gtrsim 3 \times 10^6 \Msun$ to probe subhalos that may not host luminous galaxies.
\citet{Bakels21} report that nearly half of all subhalos with $M_{\rm sub,acc} / M_{\rm host,200m} \sim 10^{-3}$ were pre-processed, and this ratio decreases for increasing subhalo mass.
When specifically analyzing subhalos \textit{on first infall}, \citet{Bakels21} showed that as many as 40 per cent of subhalos were pre-processed prior to falling into their MW-mass halo, and this fraction increases for more massive host halos.
More recently, \citet{DSouza21} used the ELVIS DMO simulations to study the times since infall of subhalos with $M_{\rm halo,peak} > 10^9 \Msun$ and how they were influenced by a massive merger ($>1:10$).
The distribution of times since infall for their surviving subhalos range $0 - 12 \Gyr$, and the satellite $\MWinfall$ are peaked toward more recent values compared to the splashback population, which were accreted earlier.
The full range of times since infall in our Figure~\ref{fig:times_mstar} (and Figures~\ref{fig:times_mhalo} and \ref{fig:dmo}) are consistent with the distribution in \citet{DSouza21}.
Although \citet{DSouza21} did not specifically focus on the first infall of subhalos into other more massive satellites/subhalos, they investigated group infall of satellites and showed that the distribution of time since infall clusters with the timing of the massive merger (and is slightly clustered with lower-mass mergers, $>1:15$), with many subhalos becoming satellites of the massive merger $< 2.5 \Gyr$ before it first crossed the MW-mass host radius.

\citet{Bakels21} showed that after first infall, subhalos generally lose orbital energy and reach apocentres that are $\approx 0.8 \times$ their turn-around radius, $r_{\rm ta}$, and all subsequent apocentres are typically comparable in distance.
On the extreme ends, some subhalos gained or lost orbital energy and thus, reached larger or smaller subsequent apocentres, respectively, analogous to our satellites with growing pericentres.
Regarding the subhalos that deviate strongly in their first apocentres and $r_{\rm ta}$, \citet{Bakels21} found that nearly $\approx 2/3$ of the satellites with first apocentres $\gtrsim 3 r_{\rm ta}$, and $\approx 80$ per cent of the satellites that only reached $\lesssim 1/4 r_{\rm ta}$, were pre-processed.
Roughly $1/3$ of the satellites with growing pericentres in our sample were pre-processed before falling into the MW, but they may also orbit outside of the more massive halo before falling into the MW halo.
Thus, it is unlikely that pre-processing is the only driving factor in the origin and orbital evolution of satellites with growing pericentres.

Both \citet{Panithanpaisal21} and \citet{Shipp22} used the same 13 MW-mass galaxies in the FIRE-2 simulations that we use here to investigate stellar stream properties.
Stellar streams form via disrupted low-mass galaxies or star clusters, however, before they completely disrupt, because they stretch throughout the halo we learn something about their initial orbits.
\citet{Shipp22} find that systems with smaller pericentres are more likely to form streams, and that the distribution of pericentres in the simulated streams are slightly smaller than the dwarf galaxies in our work.
Furthermore, the authors suggest that not only are there differences in the orbital properties of present-day satellites and stellar streams, the orbits of streams with fully or partially disrupted progenitors differ as well, highlighting the complex evolution of low-mass stellar systems.


Finally, \citet{DSouza22} explored uncertainties associated with orbit modeling using the ELVIS DMO simulations.
They suggested that using simple parametric models for the MW-mass host (and recently accreted LMC-like galaxy) result in errors that are comparable to the 30 per cent uncertainty in the halo mass of the MW.
They also extensively studied the errors associated with modeling the potential of a recently accreted LMC-like galaxy, the initial conditions of the satellites, and the mass evolution of the MW-mass halo, and they show that each comes with errors comparable to or less than the uncertainties in using simple parametric potentials.

Consistent with works like \citet{DSouza22}, our results highlight complications and limitations with idealized orbit modeling in a static, non-cosmological MW halo potential; most importantly, our results refute any expectation that the orbits of satellite galaxies always, or even generally, shrink over time.
In Santistevan et al., in prep., we will use our simulations to pursue orbit modeling of individual satellite histories to compare with idealized orbit modeling in a static host potential.



\section*{Acknowledgements} %
We greatly appreciate discussions with Nora Shipp and Pratik Gandhi throughout the development of this paper.
We are also thankful for interesting discussion with Andrey Kravtsov on the stellar mass - halo mass relation.
IBS received support from NASA, through FINESST grant 80NSSC21K1845.
AW received support from: NSF via CAREER award AST-2045928 and grant AST-2107772; NASA ATP grant 80NSSC20K0513; and HST grants GO-14734, AR-15809, GO-15902, GO-16273 from STScI.
JS was supported by an NSF Astronomy and Astrophysics Postdoctoral Fellowship under award AST-2102729.
RES gratefully acknowledges support from NASA grant 19-ATP19-0068, from the Research Corporation through the Scialog Fellows program on Time Domain Astronomy, from NSF grant AST-2007232, and from HST-AR-15809 from the Space Telescope Science Institute (STScI), which is operated by AURA, Inc., under NASA contract NAS5-26555. 
We ran simulations using: XSEDE, supported by NSF grant ACI-1548562; Blue Waters, supported by the NSF; Frontera allocations AST21010 and AST20016, supported by the NSF and TACC; Pleiades, via the NASA HEC program through the NAS Division at Ames Research Center.

\section*{Data Availability} %
The python code that we used to analyze these data is available at \url{https://bitbucket.org/isantis/orbit\_analysis}, which uses the publicly available packages \url{https://bitbucket.org/awetzel/gizmo\_analysis}, \url{https://bitbucket.org/awetzel/halo\_analysis}, and \url{https://bitbucket.org/awetzel/utilities}.
The FIRE-2 simulations are publicly available \citep{Wetzel22} at \url{http://flathub.flatironinstitute.org/fire}.
Additional FIRE simulation data is available at \url{https://fire.northwestern.edu/data}.
A public version of the GIZMO code is available at \url{http://www.tapir.caltech.edu/~phopkins/Site/GIZMO.html}.
Finally, data values in each figure are available at \url{https://ibsantistevan.wixsite.com/mysite/publications}.



\bibliographystyle{mnras}
\bibliography{paper}


\appendix

\section{Trends with Peak Halo Mass} %
\label{app:mhalo}                    %

In Section~\ref{sec:results}, we investigated trends of satellite orbital dynamics and histories as a function of satellite $\Mstar$, which is the mass most directly observable.
However, from Figure~\ref{fig:mstar_v_mhalo}, the DM (sub)halo mass of a satellite is $10^2 - 10^4 \times$ larger, so it is the one most important for dynamics.
Here we investigate the same trends but as a function of a satellite's peak halo mass, $\Mhp$.

We select \textit{all} subhalos with $\Mhp > 10^8 \Msun$, which includes both luminous and dark subhalos (with no stars).
Thus, given the extrapolated abundance matching relations of \citet{Moster13, GarrisonKimmel17_smhm, Behroozi20} in comparison to our stellar mass selected sample in Figure~\ref{fig:mstar_v_mhalo}, this includes lower-mass subhalos that likely would host ultra-faint galaxies whose stellar masses our baryonic simulations do not natively resolve.
For reference, the fraction of satellites in each mass bin that are luminous in our simulations is: 1 per cent for $\Mhp=10^{8-8.5}\Msun$, 14 per cent for $\Mhp=10^{8.5-9}\Msun$, 60 per cent for $\Mhp=10^{9-9.5}\Msun$, 92 per cent for $\Mhp=10^{9.5-10}\Msun$, and 100 per cent above $\Mhp>10^{10}\Msun$.
Compared to our stellar mass selection, this increases our satellite sample size by a factor of $\approx 8.5$.

\begin{figure}
\centering
\begin{tabular}{c}
\includegraphics[width=0.94\linewidth]{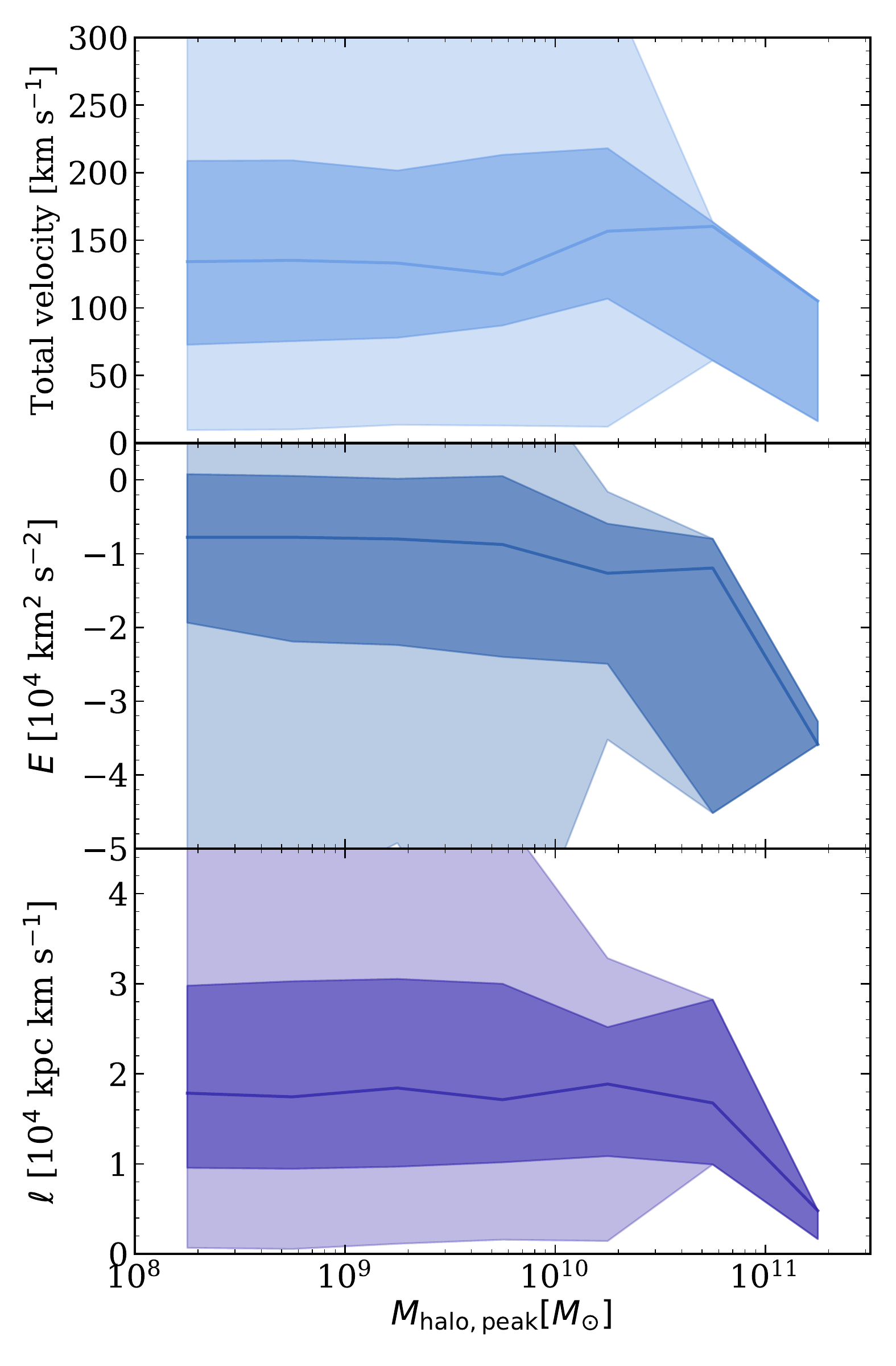}
\end{tabular}
\vspace{-3 mm}
\caption{
Similar to Figure~\ref{fig:dynamics}, the orbital dynamics of satellite galaxies at $z = 0$, here versus their peak halo mass, $\Mhp$, for all satellites (luminous and dark).
\textbf{Top}: Median orbital total velocity is nearly constant with $\Mhp$ at $\approx 135 \kmsi$ and 68th percentile ranging from $71 - 188 \kmsi$.
\textbf{Middle}: Median specific orbital total energy, $E$, is nearly constant at $-0.9 \times 10^4\kmsis$ with a 68th percentile range of $-2.6$ to $-0.2 \times 10^4\kmsis$.
Notably, the dependence on $\Mhp$ is event flatter than with $\Mstar$ in Figure~\ref{fig:dynamics}, though these quantities all decline rapidly at $\Mhp \gtrsim 5 \times 10^{10} \Msun$, likely because of the increasing importance of dynamical friction for sufficiently massive satellites.
\textbf{Bottom}: Median orbital specific angular momentum, $\ell$, is nearly constant at $\ell \approx 1.6 \times 10^4 \kpc\kmsi$ with a 68th percentile range of $0.9 - 2.6 \times 10^4 \kpc\kmsi$.
}
\label{fig:mhalo_corr_dyn}
\end{figure}

\begin{figure}
\centering
\begin{tabular}{c @{\hspace{-1ex}} c}
\includegraphics[width=0.94\linewidth]{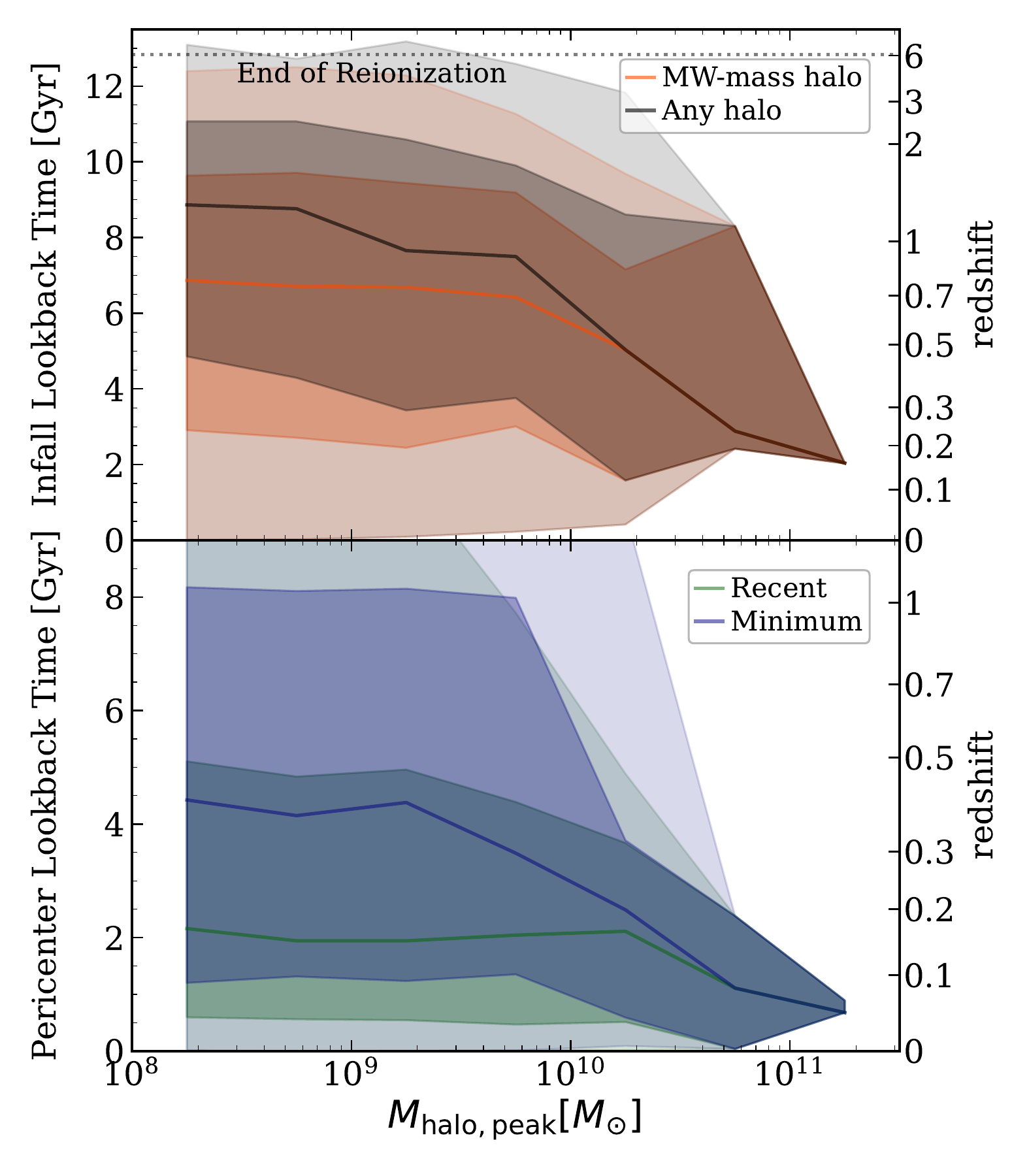}
\end{tabular}
\vspace{-3 mm}
\caption{
Similar to Figure~\ref{fig:times_mstar}, now as a function of peak halo mass, $\Mhp$, for all satellites (both luminous and dark) at $z = 0$, showing stronger and smoother trend with $\Mhp$.
\textbf{Top}: Similar with the trends with $\Mstar$, lower-mass satellites fell into any more massive halo, or their MW-mass halo, earlier than higher-mass satellites, and satellites with $\Mhp \lesssim 10^{10} \Msun$ typically fell into their MW-mass halo $\sim 1 - 2 \Gyr$ after falling into any more massive halo.
As the full distribution shows, no halos in our sample were satellites of their MW-mass hosts during the epoch of reionization ($z\gtrsim6$), even down to the ultra-faint regime, $\Mhp\approx10^8\Msun$, and less than 1 per cent of satellites were members of another more massive halo during that time.
\textbf{Bottom}: As in Figure~\ref{fig:times_mstar}, the lookback times of both the minimum and most recent pericentres occur earlier for satellites at lower $\Mhp$.
The median $\tperimin$ decreases from $\approx 4.5 \Gyr$ ago for lower-mass satellites to $\approx 0.75 \Gyr$ ago at our highest masses.
By contrast, the median $\tperirec$ is much flatter with $\Mhp$.
}
\label{fig:times_mhalo}
\end{figure}

\begin{figure}
\centering
\begin{tabular}{c}
    \includegraphics[width=0.94\linewidth]{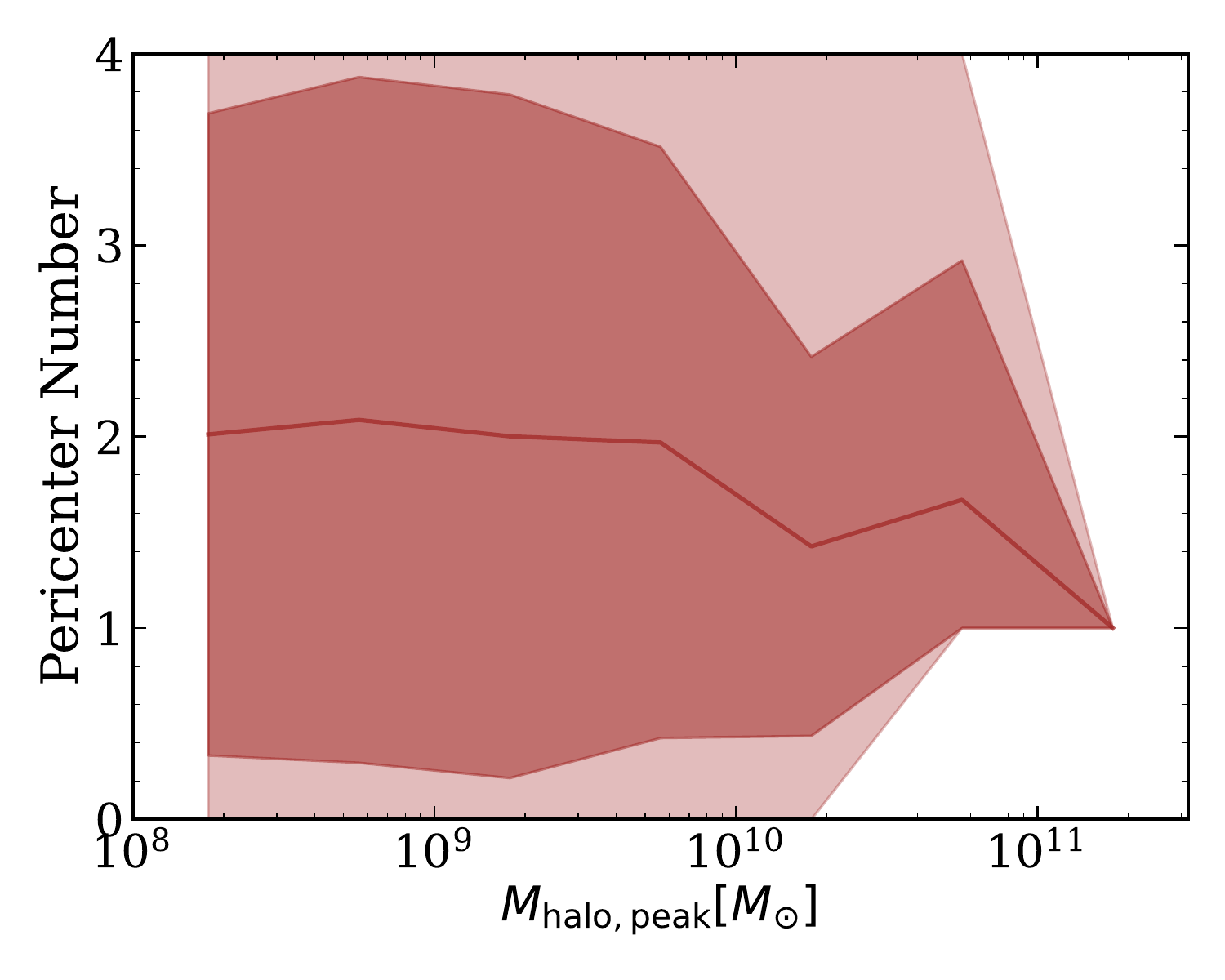}
\end{tabular}
\vspace{-3 mm}
\caption{
Similar Figure~\ref{fig:peri_dn} (top right), now showing the number of pericentric passages about the MW-mass host versus peak halo mass, $\Mhp$.
Now the trend with mass slightly weaker, decreasing from 2 to 1 with increasing $\Mhp$, because lower-mass satellites fell into the MW-mass halo earlier and on smaller orbits and do not experience as strong of dynamical friction.
The full distribution (clipped for visual clarity) is $0 - 12$ pericentres at $\Mhp \approx 10^{8.5} \Msun$.
}
\label{fig:nperi_vs_mhalo}
\end{figure}

Because (sub)halo mass is the most relevant dynamically, but a satellite with a given $\Mhp$ hosts a range of stellar masses given the scatter in the SMHM relation, trends with $\Mstar$ tend to be noisier.
Figures~\ref{fig:mhalo_corr_dyn}, \ref{fig:times_mhalo}, and \ref{fig:nperi_vs_mhalo} all show qualitatively similar trends to those in Section~\ref{sec:results}.
In particular, the trends at low halo mass in Figure~\ref{fig:mhalo_corr_dyn} show relatively flat dependence, and at $\Mhp\gtrsim10^{10.5}\Msun$, we see a more pronounced decline given the stronger dynamical friction at these masses.
Trends with the lookback times of both infall metrics and the pericentre lookback times all qualitatively show similar results and offsets in Figure~\ref{fig:times_mhalo}, and the number of pericentric passages agrees with the stellar mass selection, though it is shifted to slightly smaller $\Nperi \approx 2$ for the smallest subhalos, compared to $\Nperi \approx 2.5$ at our lowest stellar masses.
We do not show trends of pericentre distance, given the lack of a strong dependence on $\Mhp$, but we compare $\dperimin$ for satellites in baryonic versus DMO simulations in Appendix~\ref{app:dmo}.
In summary, the trends using this halo-mass selected sample are qualitatively similar to the results presented throughout Section~\ref{sec:results}.
Furthermore, our results here imply similar trends for ultra-faint galaxies, where no halo capable of hosting an ultra-faint galaxy, $\Mhp \approx 10^8\Msun$, was a satellite of the MW-mass host halo progenitor during the epoch of reionization, $z\gtrsim6$.
Similar to the results in our stellar-mass selected sample, we also find that $<1$ per cent of the satellites in this halo-selected sample were members of a more massive halo during reionization.

\section{Baryonic versus dark-matter-only simulations} %
\label{app:dmo}                                      %

Here we compare our results from our FIRE-2 baryonic simulations against satellites in dark matter-only (DMO) simulations of the same halos, to understand the effects of baryons and contextualize previous results based on DMO simulations, given that many previous works investigated satellite orbits and infall histories in DMO simulations \citep[for example][]{Wetzel15, Bakels21, DSouza21, Robles21, Ogiya21}, which, among other things, do not model the potential from a central galaxy.
Furthermore, stellar feedback in more massive satellites can reduce their inner dark-matter densities, making them more susceptible to tidal disruption \citep[for example][]{Bullock17}.
With no tidal forces from the central galaxy and with less dense dark-matter cusps within the subhalos, tidal disruption can be much stronger in baryonic simulations.
Recent studies also have used DMO simulations with an embedded disk-like potential \citep[for example][]{Kelley19, RodriguezWimberly19, Fillingham19, Robles21}.

We compare only simulations that have DMO counterparts at all snapshots, which comprises the 7 MW-mass hosts in isolated environments (names beginning with `m12') in Table~\ref{tab:hosts}.
As in Appendix~\ref{app:mhalo}, for all simulations we select all satellites with $\Mhp > 10^8 \Msun$, which includes both luminous and dark satellites in the baryonic simulations.
In the DMO simulations, we re-normalize $\Mhp$ to account for the loss of baryons by multiplying by $1 - f_{\rm b}$, where $f_{\rm b} = \Omega_{\rm baryon} / \Omega_{\rm matter}$ is the cosmic baryon fraction.
The total number of satellites in the DMO simulations is $\approx 1.6 \times$ higher.

\begin{figure*}
\centering
\begin{tabular}{c c}
\includegraphics[width=0.475\linewidth]{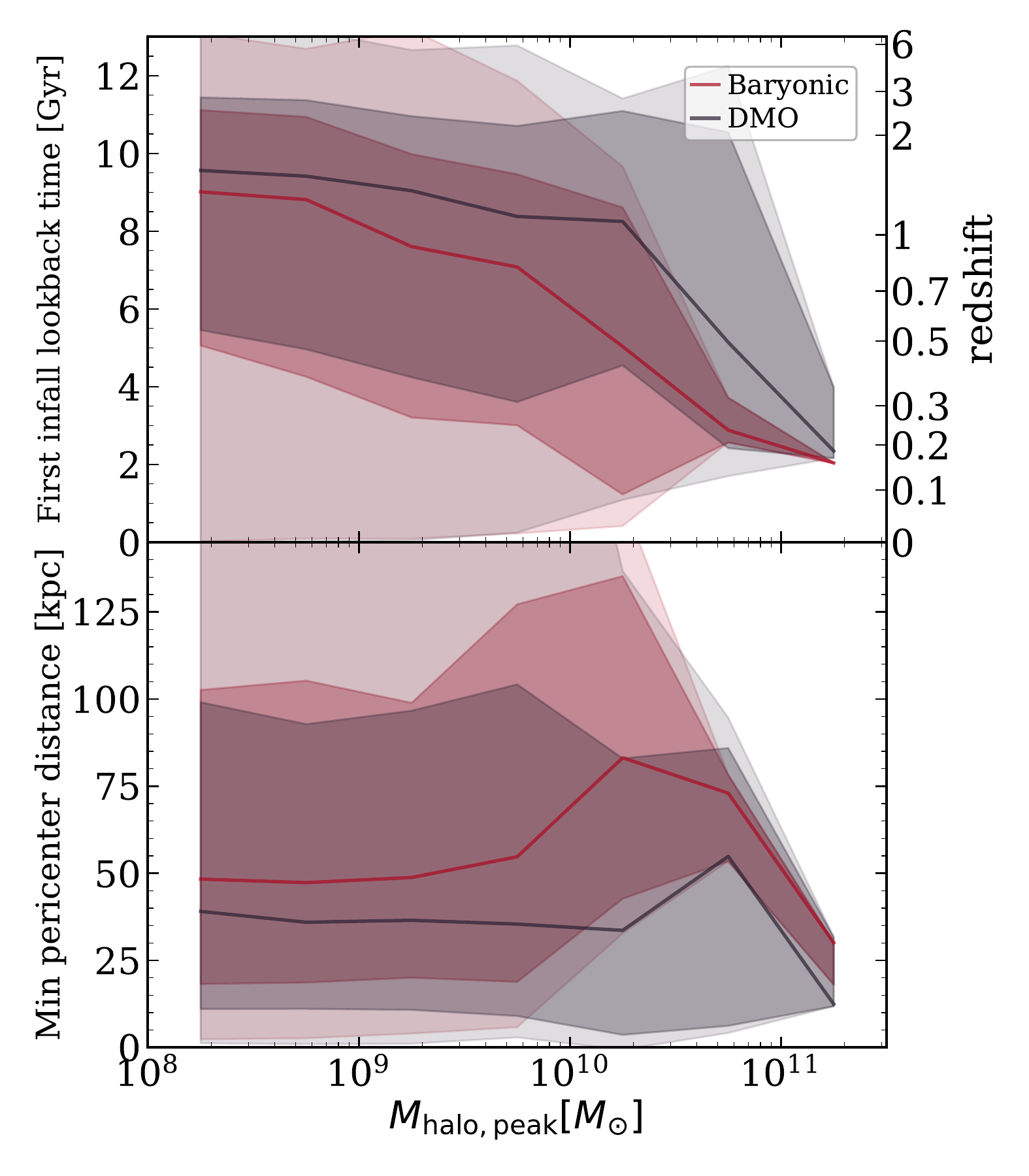}&
\includegraphics[width=0.475\linewidth]{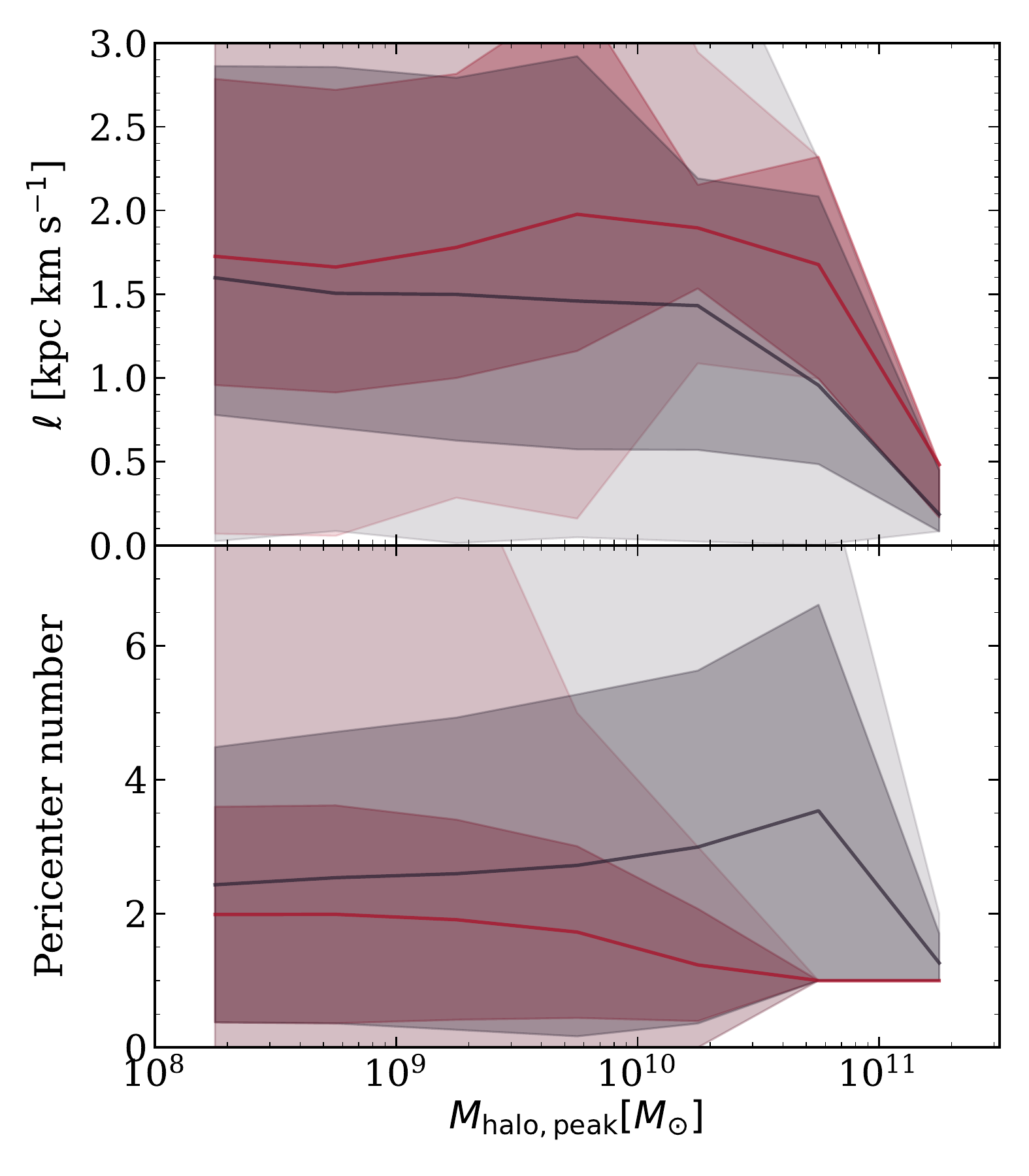}
\end{tabular}
\vspace{-4 mm}
\caption{
Comparing satellite orbital properties versus $\Mhp$ in the baryonic simulations (red) against dark matter-only (DMO) simulations (black) of the same systems, for satellites at $z = 0$ within the isolated MW-mass halos.
\textbf{Top left}: Satellites in the DMO simulations fell into a more massive halo $0.5 - 3.5 \Gyr$ earlier than in the baryonic simulations.
\textbf{Bottom left}: The median minimum pericentre distance, $\dperimin$, is smaller in DMO simulations, ranging typically from $10 - 55 \kpc$ as compared with $30 - 85 \kpc$ in the baryonic simulations.
\textbf{Top right}: The orbital specific angular momentum, $\ell$, is smaller in DMO simulations than in baryonic simulations.
\textbf{Bottom right}: The mean number of pericentric passages about the MW-mass host is smaller in the baryonic simulations than in DMO.
Also, $N_{\rm peri}$ increases slightly with satellite mass in the DMO simulations, while it decreases monotonically in the baryonic simulations.
The primary reason for these differences is that the central galaxy in the baryonic simulations induces stronger tidal stripping and disruption on satellites that orbit near it, leading to a \textit{surviving} population that fell in more recently, experienced fewer and larger-distance pericentres, and has more orbital angular momentum.
}
\label{fig:dmo}
\end{figure*}

Figure~\ref{fig:dmo} shows the lookback times of `first' infall into any other more massive halo, $\anyinfall$ (top left), specific angular momentum, $\ell$ (top right), the smallest pericentre experienced, $\dperimin$ (bottom left), and the number of pericentric passages about the MW-mass host, $\Nperi$ (bottom right), for satellites in the baryonic (red) and DMO (black) simulations.
Solid lines show the median and the dark and light shaded regions show the 68th percentile and full distribution, respectively.

Satellites in the DMO simulations do not feel the gravity of a central galaxy, so they experience weaker tidal stripping and disruption, even if they fell in early or orbit closer to the center of the halo.
Thus, the (surviving) satellites in the DMO simulations generally fell in $0.5 - 3 \Gyr$ earlier than in the baryonic simulations.
As a result of the surviving population falling in earlier, satellites in DMO simulations also orbit at smaller distances; they were able to orbit closer to the center of the MW-mass halo without becoming tidally disrupted, as the bottom left panel shows.
Furthermore, surviving satellites have lower $\ell$ in DMO simulations, given that satellites with smaller $\ell$ in the baryonic simulations are likely to be tidally disrupted \citep[for example][]{GarrisonKimmel17}.
Finally, because satellites in DMO simulations fell in earlier and orbit at smaller distances, they completed more pericentric passages (bottom right panel).
We also see a small increase in $\Nperi$ with $\Mhp$, likely because higher-mass satellites in DMO simulations in particular can survive longer than in the presence of a central galaxy.

Our results agree with \citet{GarrisonKimmel17}, who compared subhalo populations between DMO and FIRE-2 baryonic simulations using 2 of the same systems that we analyze (m12i and m12f).
They also tested the results of using a DMO simulation with an analytic galaxy potential embedded within the host halo, finding good agreement with the baryonic simulations, which implies that the most important effect in the baryonic simulations is additional gravitational effect of the MW-mass galaxy.
They showed that the number of subhalos between the different types simulations converges for subhalos that orbit farther away from the center of the MW-mass halo.
Thus, differences between DMO and baryonic simulations are largest for subhalos that orbit closer to the center of the host, where these subhalos get preferentially disrupted in the baryonic simulations, and result in a satellite population with a larger fraction on more tangential orbits, with higher specific angular momentum.

\bsp	
\label{lastpage}
\end{document}